\documentclass[useAMS,usenatbib]{mn2e}
\usepackage{natbib}
\usepackage{amsmath,amsfonts,amssymb}
\usepackage{mathptmx}
\usepackage{epsfig}
\usepackage{longtable}
\def\reff@jnl#1{{\rm#1\/}}

\def\aj{\reff@jnl{AJ}}                  % Astronomical Journal
\def\araa{\reff@jnl{ARA\&A}}            % Annual Review of Astron and Astrophys
\def\apj{\reff@jnl{ApJ}}                        % Astrophysical Journal
\def\apjl{\reff@jnl{ApJ}}               % Astrophysical Journal, Letters
\def\apjs{\reff@jnl{ApJS}}              % Astrophysical Journal, Supplement
\def\ao{\reff@jnl{Appl.Optics}}         % Applied Optics
\def\apss{\reff@jnl{Ap\&SS}}            % Astrophysics and Space Science
\def\aap{\reff@jnl{A\&A}}               % Astronomy and Astrophysics
\def\aapr{\reff@jnl{A\&A~Rev.}}         % Astronomy and Astrophysics Reviews
\def\aaps{\reff@jnl{A\&AS}}             % Astronomy and Astrophysics, Supplement
\def\azh{\reff@jnl{AZh}}                        % Astronomicheskii Zhurnal
\def\baas{\reff@jnl{BAAS}}              % Bulletin of the AAS
\def\jrasc{\reff@jnl{JRASC}}            % Journal of the RAS of Canada
\def\memras{\reff@jnl{MmRAS}}           % Memoirs of the RAS
\def\mnras{\reff@jnl{MNRAS}}            % Monthly Notices of the RAS
\def\pra{\reff@jnl{Phys. Rev. A}}         % Physical Review A: General Physics
\def\prb{\reff@jnl{Phys. Rev. B}}         % Physical Review B: Solid State
\def\prc{\reff@jnl{Phys. Rev. C}}         % Physical Review C
\def\prd{\reff@jnl{Phys. Rev. D}}         % Physical Review D
\def\prl{\reff@jnl{Phys. Rev. Lett}}      % Physical Review Letter
\def\pasp{\reff@jnl{PASP}}              % Publications of the ASP
\def\pasj{\reff@jnl{PASJ}}              % Publications of the ASJ
\def\qjras{\reff@jnl{QJRAS}}            % Quarterly Journal of the RAS
\def\skytel{\reff@jnl{S\&T}}            % Sky and Telescope
\def\solphys{\reff@jnl{Solar~Phys.}}    % Solar Physics
\def\sovast{\reff@jnl{Soviet~Ast.}}     % Soviet Astronomy
\def\ssr{\reff@jnl{Space~Sci.Rev.}}     % Space Science Reviews
\def\zap{\reff@jnl{ZAp}}                        % Zeitschrift fuer Astrophysik
\def\nat{\reff@jnl{Nature}}             % Nature 

\def\p#1by#2{{\partial{#1} \over \partial{#2}}}
\def\pp#1by#2#3{{\partial^2{#1} \over \partial{#2}\partial{#3}}}
\def\d#1by#2{{{\rm d}{#1} \over {\rm d}{#2}}}
\def\dd#1by#2#3{{{\rm d}^2{#1} \over {\rm d}{#2}{\rm d}{#3}}}

\title[AMI observations of Dark Nebulae]
{AMI observations of Lynds Dark Nebulae:
  further evidence for anomalous cm-wave emission\thanks{We request that any reference to this paper cites ``AMI Consortium: Scaife et~al. 2009"}}

\label{firstpage}
\author[Scaife et~al.]
{AMI Consortium: 
 Anna M. M. Scaife $\thanks{E-mail: 
as595@mrao.cam.ac.uk}$, 
 Natasha Hurley-Walker,
 David A. Green,
 \newauthor
 Matthew L. Davies, 
 Thomas M. O. Franzen, 
 Keith J. B. Grainge, 
 Michael P. Hobson,
\newauthor
 Anthony N. Lasenby,
 Guy G. Pooley,
 Carmen Rodr{\'i}guez-Gonz{\'a}lvez,
\newauthor
 Richard D. E. Saunders,
 Paul F. Scott, 
 Timothy W. Shimwell,
 David J. Titterington,
\newauthor
 Elizabeth M. Waldram \&
 Jonathan T. L. Zwart\\
 \vspace{0.03in}\\
 Astrophysics Group, Cavendish Laboratory, 19 J. J. Thomson Ave.,
Cambridge CB3 0HE\\
}

\date{Accepted ---; received ---; in original form \today}
\pagerange{\pageref{firstpage}--\pageref{lastpage}}
\pubyear{2005}

\begin{document}
%\label{firstpage}
\maketitle

\begin{abstract}
Observations at 14.2 to 17.9 GHz made with the AMI Small Array
  towards fourteen Lynds Dark Nebulae with a resolution of $\approx 2'$
  are reported. These sources are selected from the SCUBA 
  observations of Visser et~al. (2001) as small angular diameter
  clouds well matched to the synthesized beam of the AMI Small Array. Comparison of
  the AMI observations with radio observations at 
  lower frequencies with matched $uv$-plane coverage is made, in
  order to search for any anomalous excess emission which can be
  attributed to spinning dust. Possible emission from spinning dust
  is identified as a source within a 2\,arcmin radius of the Scuba
  position of the Lynds dark nebula, exhibiting an excess with respect
  to lower frequency radio emission. 
  We find five sources which show a possible spinning dust component
  in their spectra. These sources have rising spectral indices in the
  frequency range
  {14.2}--{17.9}\,GHz with $\alpha_{14.2}^{17.9} =
  -0.7\pm0.7$ to $-2.9\pm0.4$, where $S\propto \nu^{-\alpha}$. Of
  these five one has already been reported, L1111, we report one new
  definite detection, L675 (16\,$\sigma$), and three new probable
  detections (L944, L1103 
  and L1246). The relative certainty of these detections is assessed 
  on the basis of three 
  criteria: the extent of the emission, the coincidence of the
  emission with the Scuba position and the likelihood of alternative
  explanations for the excess. Extended microwave emission makes the
  likelihood of the anomalous emission arising as a consequence of a
  radio counterpart to a protostar or a proto-planetary disk unlikely.
  We use a 2\,arcmin radius in order to be consistent with the IRAS
  identifications of dark nebulae (Parker 1988), and our third
  criterion is used in the case of L1103 where a high flux density at
  850\,$\mu$m relative to the FIR data suggests a more
  complicated emission spectrum.

\end{abstract}


\section{Introduction}

Modern theoretical models of the infrared emission from dust in the
diffuse ISM require a large proportion of starlight energy to be
reradiated by ultrasmall grains ($\approx$20\%; Draine 2003). Comparison of the
observed spectra with theory suggests that a large fraction of these small
grains must be in the form of Polycyclic Aromatic Hydrocarbons (PAHs). It is supposed that in addition to their vibrational emission, these
small grains will also rotate (Erickson 1957; Ferrara \& Detmar 1994;
Draine \& Lazarian 1998). The emission from this rotation is
found, using simple thermodynamical arguments, to be in the cm-wave
regime of the electromagnetic spectrum. The details of the rotational
excitation and damping have been examined in detail by Draine \&
Lazarian (1998) and their models have been found to be in good
argreement with observational results.

Although the anomalous emission was originally seen as a large-scale phenomenon
(Leitch et~al. 1997, see also Kogut et~al. 1996) in CMB observations it has also been seen to
occur in a number of different types of astronomical objects, such as
{\sc{Hii}} regions, dark clouds and photodissociation regions
(Casassus et~al. 2006; Dickinson et~al. 2007; Scaife et~al. 2007;
AMI Consortium: Scaife et~al. 2009). It is often 
found to be correlated with thermal dust emission, but this is not
always the case (Casassus et~al. 2008). A growing number of observations have detected dust-correlated cm-wave 
emission which is inconsistent with free--free emission. However,
these few observations are not yet sufficient to make any definite
observational statement about the enviromental factors which influence
whether emission from spinning grains will be seen towards a certain
object. We hope that the observations presented here will contribute
significantly to improving this knowledge.

In this paper we have imaged fourteen dark molecular clouds. Our sample is selected from the submillimetre survey of dark
molecular clouds performed using the Submillimetre Common-User
Bolometre Array (SCUBA) on the James Clerk Maxwell Telescope (JCMT) by
Visser, Richer \& Chandler (2001). Of these
fourteen only three are currently known to contain protostars (L944,
L1014 and L1246; Visser
et~al. 2001; Young et~al. 2004), with the remainder thought to be examples of pre-stellar
cores. 

The dark molecular clouds in this sample were taken from the Lynds dark
nebula catalogue (Lynds 1962) of 1802 optically selected dark clouds
from the POSS plates. The SCUBA jiggle mapping
function used to observe the sample restricted the field of view to
2.3 arcmin diameter. This constraint led to a selection criterion of a
maximum semi-minor axis of $<1.1$\,arcmin for each cloud, which was
determined from their optical size as listed in Parker (1988). 

The Lynds catalogue is divided into opacity classes
from 1 to 6 of which 6 is the most opaque and includes 147 clouds. The
clouds observed in the SCUBA survey are all class 6 ($A_{\nu} \geq
5$\,mag; Lynds 1962) and should
therefore provide the highest column density of dust when searching for
anomalous emission from spinning grains. 
 
We have observed all clouds in the sample of Visser et~al. (2001) which fall in the
available declination range of the Arcminute Microkelvin Imager (AMI)
Small Array (SA) 
instrument in Cambridge. The small angular size of these clouds makes
them well matched to the size of the AMI SA synthesized beam. We
identify excess microwave emission from the SCUBA clouds as any source
which falls within a 2\,arcmin radius of the SCUBA position and which
shows an excess of emission relative to lower frequency radio data. In
addition, any source which satisfies these two criteria but has a
falling spectrum across the AMI band will be rejected as a spinning
dust candidate. This allows us to distinguish those sources which may
be optically thick at lower radio frequencies, but have spectra which turn
over before the AMI band.

The paper is structured as follows: Section 2 describes the telescope and the observations, Section 3
details the methods we use to compare our data to lower frequency
observations, Section 4 discusses each object individually and Section
5 discusses the sample as a whole and states our conclusions.

\section{Observations}

AMI comprises two synthesis arrays, one of ten 3.7\,m
antennas (Small Array) and one of eight 13\,m antennas (Large Array),
both sited at Lord's Bridge, Cambridge (AMI Consortium: Zwart et~al. 2008). The telescope observes in
the band 13.5--17.9\,GHz with cryostatically cooled NRAO indium-phosphide
front-end amplifiers. The overall system temperature is approximately
25\,K. The radio frequency is mixed with a 24\,GHz local oscillator,
down converting to an intermediate frequency band of
6--10.5\,GHz. Amplification, equalization, path compensation and
automatic gain control are then applied to the IF signal. The
correlator is an analogue Fourier transform spectrometer with 16
correlations formed for each baseline at path delays spaced by
26\,mm. In addition both real and imaginary correlations are formed by
use of 0$^{\circ}$ and 180$^{\circ}$ hybrids. From these, eight
0.75\,GHz bandwidth channels are synthesized. In practice, the two
lowest frequency channels (1 \& 2) are not generally used due to a low response
in this frequency range and interference from geostationary
satellites. 

Observations of fourteen dark molecular clouds from the Lynds
dark nebula catalogue were made with the AMI SA telescope during the
period 2007 January -- May. The AMI LA was not used in this
work. Here we present data for each of these
regions, which are listed in Table~\ref{tab:lclouds}. These targets were chosen
from the submillimetre survey of Visser et~al. (2001) as small
molecular clouds, well matched to the AMI synthesized beam, in a
declination range consistent with the capabilities of the AMI at a
latitude of $52^{\circ}$ N. Each pointing was observed at least twice
for at least eight hours, with the exception of L1185 which was
observed only once due to poor weather.

\begin{table*}
\centering
\caption{AMI Lynds cloud sample. Column [1] gives the source name,
  Column [2] the Right Ascension, Column [3] the declination, Column
  [4] lists the r.m.s. specific intensity outside the primary beam for
  each map, Column [5] lists the optical Semi-major Axis of the cloud,
  Column [6] lists the optical Semi-minor Axis of the cloud, Column [7] lists the major axis of the synthesized beam,
  Column [8] lists the minor axis of the synthesized beam and Column
  [9] lists the complex associated with each dark cloud.\vspace{0.2cm}\label{tab:lclouds}} 
\begin{tabular}{ccccccccc}
\hline\hline
Name & RA$^{1}$ & Dec$^{1}$ & $\sigma_{\rm{rms}}$ &Semi-major
Axis$^{2}$&Semi-minor Axis$^{2}$ &$\Delta
\theta_{\rm{max}}$&$\Delta \theta_{\rm{min}}$& Associated \\
&(J2000)&(J2000)&($\frac{\rm{mJy}}{\rm{beam}}$)&(arcmin)&(arcmin)&(arcmin)&(arcmin)& complex\\
\hline
L675 & 19 23 52.6 & 11 07 39 & 0.17 &2.2&0.6&2.6&2.4&\\
L709 & 19 13 54.1 & 16 26 58 & 0.31 &1.1&1.1&2.5&2.4&\\
L860 & 20 03 18.1 & 36 01 45 & 0.21 &1.1&1.1&2.5&2.1& Cloud B\\
L917 & 20 39 55.1 & 44 08 41 & 0.27 &1.7&0.6&2.8&2.1& Cloud B\\
L944 & 21 17 40.8 & 43 18 08 & 0.27 &1.7&1.1&2.5&2.3& Cygnus Rift\\
L951 & 21 20 13.0 & 43 31 30 & 0.25 &2.2&0.6&2.5&2.2& Cygnus Rift\\
L953 & 21 21 25.5 & 43 21 04 & 0.55 &1.1&0.6&2.5&2.4& Cygnus Rift\\
L1014 & 21 24 01.7 & 49 59 26 & 0.38 &1.7&0.6&2.5&2.1& Cygnus Rift\\
L1021 & 21 21 39.3 & 50 57 20 & 0.40 &1.1&0.6&2.4&2.1& Cygnus Rift\\
L1103 & 21 42 10.2 & 56 43 44 & 0.48 &2.8&1.1&2.4&2.1&\\
L1111 & 21 40 27.1 & 57 48 10 & 0.29 &1.7&1.1&2.4&2.1&\\
L1166 & 22 05 26.2 & 59 33 38 & 0.56 &1.1&0.6&2.6&2.2&\\
L1185 & 22 29 24.5 & 59 09 23 & 1.12 &2.8&0.6&2.8&2.2&\\
L1246 & 23 25 12.9 & 63 39 30 & 0.36 &2.8&1.1&2.5&2.2& Cepheus OB3 \\
          &            &            &   &&&&   & complex\\

\hline
\end{tabular}
\begin{minipage}{14cm}
{\small\vspace{0.1cm} $^{1}$ positions adapted from Visser
  et~al. (2001)\\
$^{2}$ Parker (1988)} 
\end{minipage}
\end{table*}

Data reduction was performed using the local software tool
\textsc{reduce}. This applies
both automatic and manual flags for interference and
shadowing and hardware errors. It also applies phase and amplitude
calibrations; it then
Fourier transforms the correlator data to synthesize frequency
channels 
before output to disk in \emph{uv} FITS format suitable for imaging in
\textsc{aips}.

Flux calibration was performed using short observations of 3C286 near
the beginning and end of each run. We assumed I+Q flux densities for this source in the
AMI SA channels consistent with the frequency dependent model of Baars et al. (1977), $\simeq 3.3$\,Jy at
16\,GHz. As Baars et~al. measure I and AMI SA measures
I+Q, these flux densities 
include corrections for the polarisation of the calibrator source derived
by interpolating from VLA 5, 8 and 22\,GHz observations\footnote[1]{{\tt http://www.vla.nrao.edu/astro/calib/manual/polcal.html}}. A correction is
also made for the changing intervening air mass over the observation. From
other measurements, we find the flux calibration is accurate to better than
5 per cent (AMI Consortium: Scaife et~al. 2008; AMI Consortium: Hurley--Walker et~al. 2009).

The phase was calibrated using hourly interleaved observations of
calibrators 
selected from the Jodrell Bank VLA Survey (JVAS; Patnaik et~al. 1992). After calibration, the phase is generally stable to
$5^{\circ}$ for channels 4--7, and
$10^{\circ}$ for channels 3 and 8. The FWHM of the primary beam of the AMI SA is $\approx 20$\arcmin at
16\,GHz. Channels 1 and 2 are generally discarded due to heavy
satellite interference.

Reduced data were imaged using the AIPS data package. Both the {\sc{mem}} and
{\sc{clean}} deconvolution methods were employed, as outlined in the next
section. {\sc{clean}} deconvolution was performed using the task
{\sc{imagr}} which applies a differential primary beam correction to
the individual frequency channels to produce the combined frequency
image. {\sc{clean}} deconvolution maps were made from both the
combined channel set and for individual channels. The full set of
{\sc{clean}}ed AMI maps is shown in Appendix A. With the exception
of Fig.~\ref{fig:l675both} all of the images presented in this paper were made
using {\sc{clean}} deconvolution and are not primary beam
corrected. MEM deconvolution is also used on occasion in this paper,
for reasons described in Section~\ref{sec:analysis}, and is
implemented through the {\sc{aips}} task {\sc{vtess}}. The broad
spectral coverage of AMI allows a representation of the spectrum
between 14.2 and 17.9\,GHz to be made independently of other
telescopes and in what
follows we use the convention: $S\propto \nu^{-\alpha}$, where $S$ is
flux density, $\nu$ is frequency and $\alpha$ is the spectral index. 

\section{Multi-frequency Analysis}
\label{sec:analysis}

The synthesized beam of AMI SA has FWHM $\approx$ 2--3 arcminutes. This
makes comparison with low frequency surveys 
difficult as the GB6 survey at 4.85\,GHz (Gregory et~al. 1996) and the
Effelsberg telescope surveys at 2.7\,GHz (F${\ddot{\rm{u}}}$rst et~al. 1990) and 1.4\,GHz (Reich
et~al. 1997) have larger beams. Moreover, flux losses are important in the context of the
observations presented here as all of the sample of dark clouds lie
within the Galactic plane, where the emission on different scales is
complex, and will be represented differently by telescopes with
different effective 
{\it uv}-coverage. To constrain the spectral behaviour of these
sources we employ two {\it{uv}} matching strategies which are outlined below. Similarly the information provided on these regions at 100~$\mu$m by the IRAS satellite and its recently re-processed versions (Miville-Desch$\hat{\rm{e}}$nes \& Lagache 2005; Schlegel, Finkbeiner \& Davis 1998) are at a resolution of $\approx$ 4~arcmin, larger than that of AMI. Consequently we rely on the finer resolution results of the SCUBA observations to constrain the dust properties of the clouds.

In order to obtain spectra for these sources, direct comparison with
lower frequency surveys is not possible. We must therefore perform
some operation on the data in order to match the observed angular
scales. In the case that $b_{\rm{SD}} \ll b_0$, where $b_{\rm{SD}}$ is
the beam of the single dish telescope and $b_0$ is the synthesized
beam of the interferometer, the two sets of
visibilities and their resulting maps can be compared directly with
little loss of information, since the Fourier transform of the single
dish beam, $\tilde{b}_{\rm{SD}}(u,v) \approx 1$
 over the $(u,v)$ range of interest.
 However, this is not true in the case that $b_{\rm{SD}}$ approaches or is greater than $b_0$.

For nine of our fourteen Lynds clouds lower frequency radio
observations are available at 1.4\,GHz from the Canadian Galactic
Plane Survey (Taylor et~al. 2003; CGPS hereinafter). The CGPS is a
combination of fully sampled interferometric data and single dish
data, which provides a total power measurement of the sky with a beam
of $60'' \ll 
b_{\rm{AMI}} \simeq 150''$ (depending on declination). For the
nine clouds covered by the CGPS we can make the approximation that
$\tilde{b}_{\rm{SD}}(u,v) \approx 1$ over the {\it uv} range of interest
and compare \emph{uv}-sampled CGPS data directly to our measured visibilities
from AMI. However, in the case of the remaining five clouds we are
limited by the resolution of the available data and can only make a map
plane comparison, after convolving the clean AMI map to a common
resolution. This leads to two distinct analysis paths.  

The details of these data reduction methods 
are of importance when comparing flux densities from different
instruments. This is especially the case when using
interferometers. We include them here to emphasise the point that
when we compare AMI observations to those at lower radio frequencies we are
comparing matched angular scales, unless stated otherwise.

\subsection{Image reconstruction}

To simulate the visibilities we
take the single dish map and multiply by the primary beam of AMI, we
then FFT the modulated map to obtain the data in the form of
visibilities. These visibilities are sampled according to the
measured visibilities in the matching AMI observation. The sets of
sampled visibilities are each imaged and
deconvolved 
separately to obtain our final
simulated maps.

\noindent
{\textbf{Case 1:}} CGPS data, $b_{\rm{SD}} \ll b_0$: We use data from the CGPS archive at 1.4\,GHz for simulations. These
data have a resolution of $b_{\rm{SD,FWHM}}=60''$. We can examine the
correlation between the two data sets in both the map plane and the
{\it uv} plane.

\noindent
    {\textbf{Case 2:}} non-CGPS data, $b_{\rm{SD}} \gtrsim b_0$: For
    example, data from the Effelsberg 100\,m telescope survey at
    2.7\,GHz. These  
data have a resolution of $b_{\rm{SD,FWHM}}=4\farcm3$. In this case we
cannot directly compare the visibilities in the {\it{uv}} plane, since
the data is multiplied by the Fourier transform of the convolving
beam, $b_{\rm{SD}}$. In the map plane, however, we can recover usable
datasets for comparison when the MEM method of deconvolution is
applied.

In brief, when deconvolving interferometric data using the {\sc{clean}} algorithm a
recovered sky will have the form:
\begin{equation}
I(l,m)_{\rm{recovered}} = [I(l,m)\times B(l,m)]\ast b_0(l,m)
\end{equation}
\noindent
where $I(l,m)$ is the true sky distribution, $B(l,m)$ is the primary
beam response and $b_0(l,m)$ is the clean beam. In the case where we
sample single dish data our input intensity distribution is no longer
the true sky 
but instead a model sky, which is the 
convolution of the true sky with the single dish beam, $I'(l,m) =
I(l,m)\ast b_{\rm{SD}}$. Since this
convolution is non-associative with the multiplication by the primary
beam response, the two maps cannot be quantitatively compared.

However, if instead we choose to deconvolve using an MEM method, such
as the AIPS task VTESS, the primary beam response is accounted for in the
fitting. MEM lets us recover:
\begin{eqnarray}
\nonumber I(l,m)_{\rm{recovered}} &=& I'(l,m)\ast b_0(l,m)\\
&=& I(l,m)\ast b_{\rm{SD}}\ast b_0(l,m).
\end{eqnarray}
\noindent
Since convolution is associative with itself we can then convolve our
own AMI data to match our sampled data. We do this by deconvolving the
AMI data using VTESS to recover $I(l,m)\ast b_0(l,m)$ and convolving
this recovered map with the single dish beam $b_{\rm{SD}}$. This will
result in our 
comparison maps, both the sampled data and the convolved AMI data, having a clean beam of $b(l,m) = \sqrt{b_0(l,m)^2 +
  b_{\rm{SD}}(l,m)^2}$. We note that this process will degrade the original
AMI data, however it results in maps which have identical resolution
and contain the same angular scales. 

\subsection{Flux extraction}
\label{sec:poly}

The morphology of the sources in these observations is often not well
described by a Gaussian fit. We therefore 
estimate their flux densities by removing  a tilted plane fitted 
to the local background and
integrating the remaining flux, see for example Green (2007). We do
this by drawing a polygon around each source and fitting a tilted
plane to the pixels around the edge of the polygon. Where an edge of
the polygon crosses a region confused by another 
source the background is subjective and we omit this edge from the
fitting. Example polygons are shown in e.g. Fig.~\ref{fig:l675both} with 
omitted edges shown as dashed lines. Where we believe there may be
some confusion as to the aperture selected in what follows we have
depicted an example aperture in the same way, see the figure captions for
details. Since this method is dependent to some
degree on the aperture selected around the source we perform five fits
for each object changing the aperture slightly each time, and take the flux density as the average of these
fits. The vertices of these apertures are listed in Appendix A2. We denote the variance of these fits as $\sigma_{\rm{fit}}^2$
and include it in the final error on the calculated flux density as:
\begin{equation}
\sigma = \sqrt{\sigma_{\rm{rms}}^2+(0.05S)^2+\sigma_{\rm{fit}}^2}.
\end{equation} 
Here we represent the r.m.s. fluctuations outside the primary beam
measured from the map as $ \sigma_{\rm{rms}}$ and  the flux density of the
source as $S$. This calculation assumes a conservative 5 percent error
on the flux calibration. 

We note that images made using CLEAN are often systematically
different to those made using MEM through VTESS. The major, although
not only, difference
arises due to the positive definite nature of the model assumed in
VTESS. This has the effect of raising the background level in images
reconstructed by VTESS relative to those made using
CLEAN. Consequently it is not recommended to use these MEM
reconstructions for comparison with images restored with CLEAN. Our
method of flux extraction acts to mitigate this effect through
subtracting a tilted plane. However, we note that we have not used MEM
images for quantitative comparison except in the case of L675 where we
have compared like with like. In later sections we have sometimes
included a flux density recovered using an MEM map to enhance a flux
spectrum, but this is purely for illustrative purposes and will be
explicitly commented upon in the text. We reiterate: Case (1) employs
CLEAN deconvolution, Case (2) employs MEM deconvolution.

\subsection{Sub-mm data}
\label{sec:greybody}

The SCUBA data of Visser et~al. (2001;2002) in conjunction with
IRAS data available from the literature allows us to place constraints
on the thermal dust spectrum of these objects and their dust
temperature, $T_d$.

Following Andre, Ward-Thompson \& Barsony (1993) we use a modified greybody
spectrum of the form 
\begin{equation}
S_{\nu} = B_{\nu}(T_d)\left(1-{\rm{e}}^{-\tau_{\nu}}\right)\Omega_{S,\nu}
\end{equation}
\noindent
to fit these data, where $B_{\nu}(T_d)$ is the Planck spectrum for a
temperature $T_d$ at a frequency $\nu$. $\tau_{\nu}$ is the optical
depth of the cloud, generally assumed to be proportional to
$\nu^{1.5}$, and $\Omega_{S,\nu}$ is the extent of the source at a given
frequency. The inclusion of $\Omega_{S,\nu}$ allows us to account for
the different source sizes seen by IRAS and SCUBA.

\section{Identification of radio counterparts}

In this section we describe the radio emission seen towards the Lynds
dark nebulae listed in Table~\ref{tab:lclouds} between 14.2 and
17.9\,GHz. Sources are listed in Table~\ref{tab:list}. In what follows
we identify potential excess microwave emission from the SCUBA clouds
as any source 
which falls within a 2\,arcmin radius of the SCUBA position and which
shows an excess of emission relative to lower frequency radio data. In
addition, any source which satisfies these two criteria but has a
falling spectrum across the AMI band will be rejected as a spinning
dust candidate. This allows us to distinguish those sources which may
be optically thick at lower radio frequencies, but have spectra which turn
over before the AMI band. All errors are quoted to 1\,$\sigma$.
 
\begin{figure}
\centerline{\includegraphics[width=8.cm,angle=0]{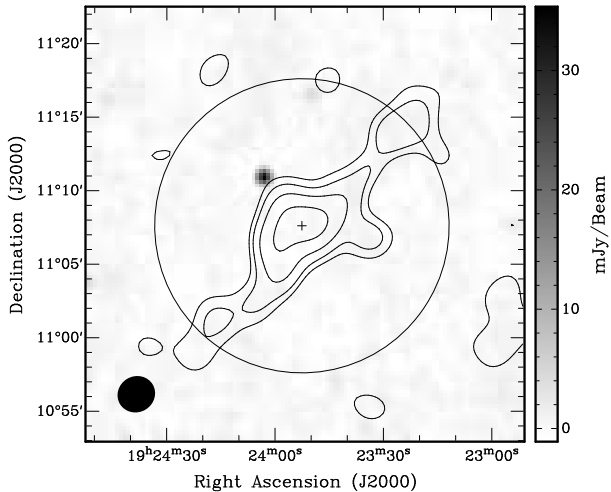}}
\caption{L675: CLEANed combined channel AMI data at 16\,GHz are shown
  as contours. Contours are at 3, 6, 12, 18 $\sigma_{\rm{rms}}$
  etc. NVSS data are shown as greyscale. The
  AMI primary beam FWHM is shown as a solid circle and the pointing centre
  as a cross. The AMI synthesized beam, $2\farcm6 \times 2\farcm4$, is shown as
  a filled ellipse in the bottom left corner. This map is not primary beam
  corrected. \label{fig:l675}
}
\end{figure}

\begin{figure*}
\centerline{\includegraphics[height=6.5cm,width=8.cm,angle=0]{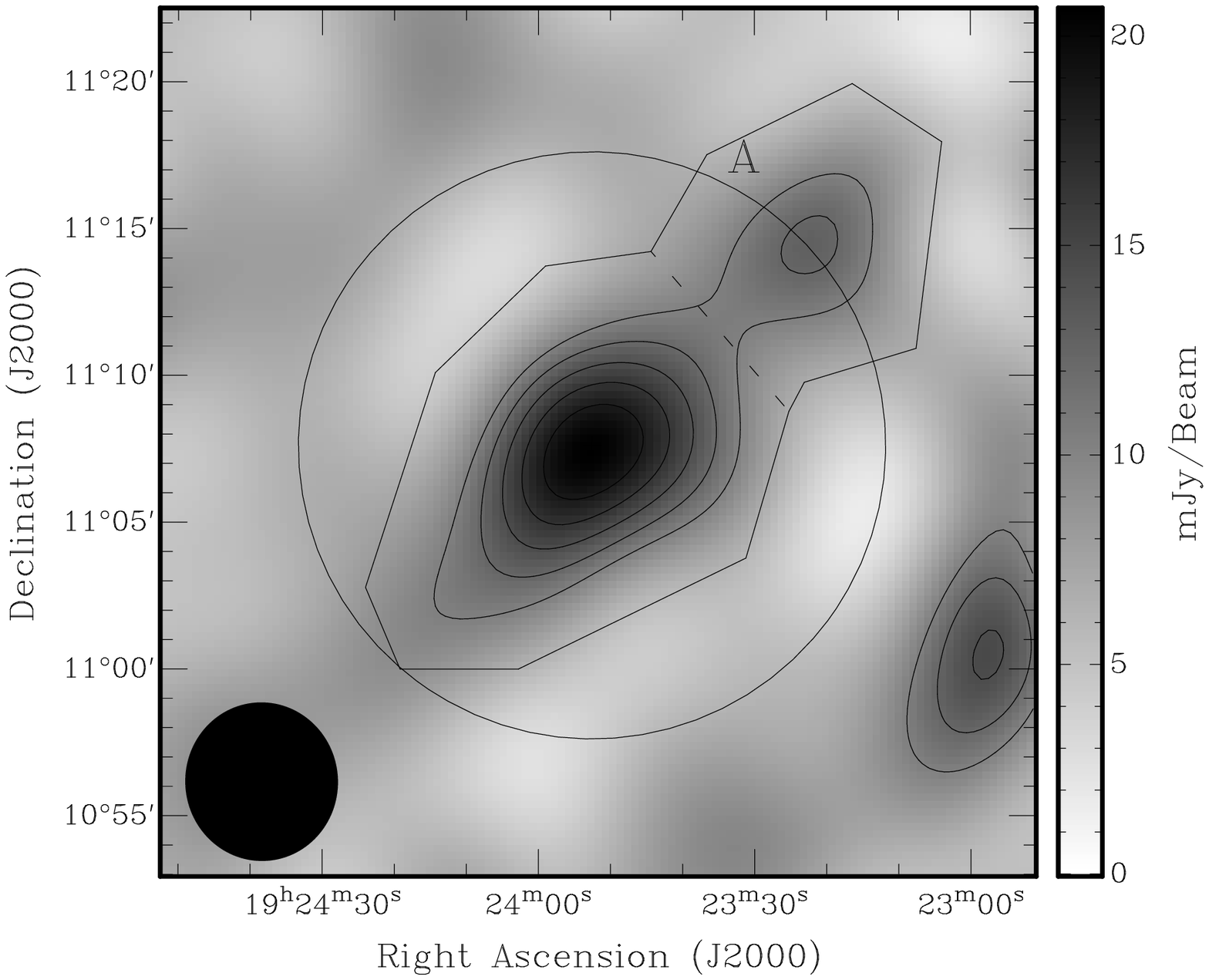}\qquad\includegraphics[height=6.5cm,width=8.cm,angle=0]{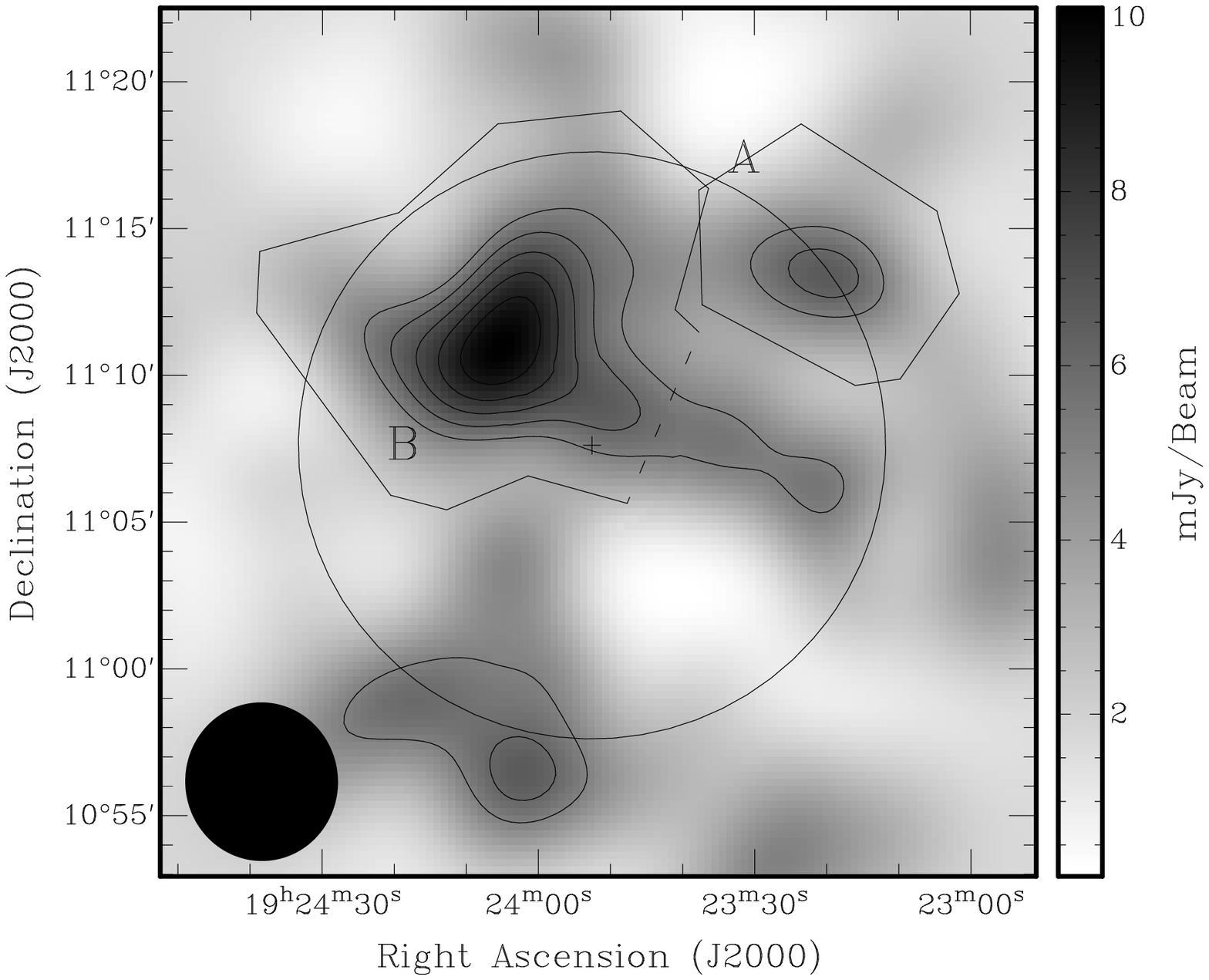}}
\centerline{(a) \hspace{8cm} (b)}
\caption{L675: (a) AMI 16\,GHz combined channel data convolved to
      mtach the resolution of Effelsberg 2.7\,GHz and (b) Effelsberg 2.7\,GHz data sampled to match the
      {\it{uv}} coverage of AMI. Both maps have a resolution of $5\farcm4\times
    5\farcm2$, shown as a filled ellipse in the bottom left
      corner, and have been deconvolved using MEM. The AMI primary
      beam is shown as a circle and the
      pointing centre as a cross. These maps are primary beam
      corrected. Contours
      are 10 percent from the 50 percent level. Example flux
      extraction apertures are shown where a dotted line indicates an
      edge not used for fitting a titled plane base-level, see text
      for details.\label{fig:l675both}}
\end{figure*}

\subsection{L675:}\label{sec:l675} L675 shows a large
radio counterpart at 16\,GHz. This region lies outside the the limits
of the CGPS data, however it is covered by NVSS at 1.4\,GHz,
Effelsberg at 2.7\,GHz and the  GB6 survey at
4.85\,GHz. These data show no counterpart to the object seen in the
higher frequency AMI map, see Fig.~\ref{fig:l675}.

Since there is no coverage of this region by the CGPS we use our
analysis Case (2) to determine the spectrum of this object. Sampling
data at 2.7\,GHz and 4.85\,GHz from the Effelsberg 100\,m telescope\footnote[1]{{\tt http://www.mpifr-bonn.mpg.de/old\_mpifr/survey.html}}
and the GB6 survey\footnote[2]{{\tt http://skyview.gsfc.nasa.gov/cgi-bin/query.pl}} respectively, and correcting for the beams as
described in Section~\ref{sec:analysis} it is immediately noticeable that the
morphology of this object is very different at 16\,GHz to that of the
lower frequencies: see Fig.~\ref{fig:l675both}. A correlation analysis of the maps shows that the
AMI and Effelsberg data have a Pearson coefficient of only 0.25, and
AMI and GB6 have a slightly higher correlation coefficient of
0.38. The correlation of GB6 and Effelsberg is 0.56. However, the
correlation between the AMI map and appropriately sampled IRIS
100\,$\mu$m data$^2$ is 0.82, demonstrating that the 16\,GHz emission follows the
infra-red emission much more closely than the low frequency radio
data. Since the  object is not detected in the lower frequency maps we
take the 
intensities at the position of the cloud at those wavelengths and regard them as upper limits on any
emission. These values are $S_{2.7} \leq 6.2$\,mJy\,beam$^{-1}$ for Effelsberg
data and $S_{4.85} \leq 3.8$\,mJy\,beam$^{-1}$ for the GB6 data. The
spectral index, $\alpha_{2.7}^{4.85}\geq 0.84$, is consistent with a
region of non-thermal emission and predicts a flux density of 1.4\,mJy
at 16\,GHz. At
the same resolution, $5\farcm4 \times 5\farcm2$, the AMI emission has a
flux density of $S_{16} = 20.2\pm1.02$\,mJy. This indicates
an excess of 16.4\,mJy (16.1\,$\sigma$) at  
16\,GHz relative to a flat spectral index from 4.85\,GHz or an excess
of 18.8\,mJy (18.4\,$\sigma$) relative to the predicted flux density from $\alpha_{2.7}^{4.85}$, assuming a
source size equivalent to the beam at lower radio frequencies. Note that the noise level in the original GB6 data at
this declination is $\approx 5$\,mJy\,beam$^{-1}$ (Gregory et~al. 1996) and in the
original Effelsberg data is $\approx 8$\,mJy\,beam$^{-1}$ (F$\ddot{\rm{u}}$rst et~al. 1990)  placing all
the structure within 
the AMI primary beam below the $2\sigma$ noise level at both frequencies.

Two further sources are evident within the primary beam of the AMI
field. Of the two point sources, Source ``A'' lies almost exactly on the FWHM point of
the beam to the north-west of L675. It appears in the six AMI channels  
and Effelsberg data with approximately the same flux density: $S_{2.7}
= 4.3$\,mJy. The best fit spectral index from these data is
$\alpha=0.00\pm0.15$ . We
identify this source as a region of optically thin thermal
emission. Note that the derived flux density of this source at
2.7\,GHz lies below the
noise level of the original Effelsberg data and consequently we treat
it as an upper limit. Source ``B'', see Fig.~\ref{fig:l675both}, appears both in the NVSS map as a compact
object and in the Effelsberg data as a more extended source, with
flux densities of $S_{1.4} = 40.0\pm1.6$\,mJy and $S_{2.7} =
13.1\pm4.9$\,mJy respectively. This indicates a steep spectrum source
and, although 
there may be a small amount of flux at 16\,GHz as indicated by the
morphology of the main object, there is no identifiable discrete
object at this frequency. 

Although there is no object associated with L675 in the IRAS point
source catalogue (hereinafter IRAS PSC) it was part of a deep IRAS
photometry catalogue (Clemens, Yuo \& Heyer 1991; hereinafter CYH91). This catalogue
yielded fluxes for L675 at all four IRAS frequencies and a size of
$\Omega_S = 39$\,arcmin$^2$, which corresponds well to the AMI
data (note that the source size from CYH91 corresponds to all flux down to
the fitted background, not a Gaussian FWHM). 

A greybody model, as described in Section~\ref{sec:greybody}, where $\tau_{\nu} \propto \nu^{1.5}$, fitted to the 100/60\,$\mu$m data
 gives a dust temperature of $T_d =
23.0$\,K. This value agrees well with that of Clemens, Yuo \& Heyer
who fit a temperature of $T_d = 23.4$\,K. The model fitted to the IRAS
data agrees with the
SCUBA data at 850\,$\mu$m (Visser et~al. 2001) assuming the smaller
 angular size of $1\farcm5$, which may be a result of the 2\,arcmin chop on the SCUBA
jiggle map. Using the source size of CYH91, this model predicts a flux from thermal dust emission at
16\,GHz of only 0.22\,$\mu$Jy.

\begin{figure}
\centerline{\includegraphics[height=8.cm,angle=-90]{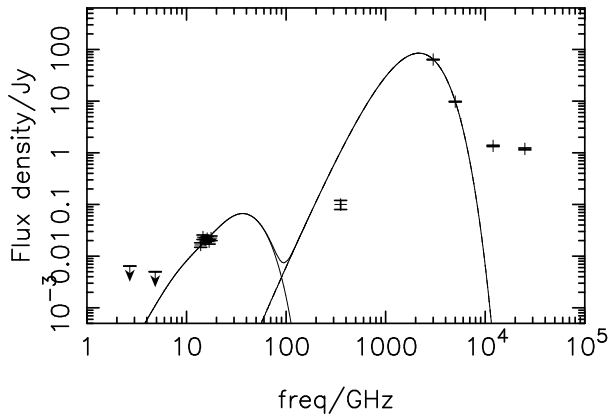}}
\caption{L675: Upper limits are shown at 2.7 and 4.85\,GHz from the
  Effelsberg 100\,m telescope and the GB6 survey respectively. These
  data are {\it{uv}} matched upper limits. Data at 14.2 to 17.9\,GHz
  from AMI are shown, as are points at 850\,$\mu$m (353\,GHz) from
  Visser et~al. (2001; 2003); and at 100, 60, 25 and 12\,$\mu$m from
  CYH91. A modified greybody spectrum with
  $\beta = 1.5$ is fitted to the 100 and 60\,$\mu$m data as
  described in the text and a MC spinning dust model from DL98 is
  fitted to the AMI data points. The combined SED is shown as a solid line.\label{fig:l675allspec}}
\end{figure}

We fit three models to the AMI data points along with the upper limits
at 2.7 and 4.85\,GHz. The first, ignoring the lower frequency
constraints, is the best fitting power law to the AMI channel data alone. The
spectral index is not well constrained: $\alpha_{\rm{AMI}} = -0.7\pm0.7$. Including
the constraint given by the upper limit at 4.85\,GHz requires $\alpha
\leq -1.15$. The third model we fit is not a power law but the
spinning dust spectrum for cold, dark molecular clouds (Drain \& Lazarian 1998;
hereinafter DL98). Using a source dimension of $\Delta \theta =
5.5$\,arcmin from the AMI map, this model fits the data with $N({\rm{H_2}}) =
(2.4\pm1.3)\times 10^{25}$\,m$^{-2}$ (where
$N({\rm{H_2})}=0.5N(\rm{H})$\footnote[1]{\tt http://www.astro.princeton.edu/$\sim$draine/dust/\\dust.mwave.spin.html}). Visser et~al. (2001) find a peak
and averaged column density of 9 and $3.4\times 10^{25}$\,m$^{-2}$ for
L675, consistent with this value. We note that the extent of L675 in
the SCUBA map is small ($\approx 1\farcm5$), if this is a consequence
of the SCUBA chop then the column density values may be
underestimated. The full spectrum for the spinning dust model, 
along with the best-fit thermal dust model, is shown in Fig.~\ref{fig:l675allspec}.

One further possibility for L675 may be that we are seeing an
ultracompact {\sc{Hii}} region. A spectral index of $\alpha = -1.15$
would be broadly consistent with that expected for such a
region. However, to have a turn over frequency greater than 20\,GHz
 (assuming an electron temperature of $\approx$8000\,K) would require an emission measure in excess of
$1.2\times10^9$\,pc\,cm$^{-6}$. An emission measure of this magnitude would
imply a mass for the {\sc{Hii}} region of hundreds of solar
masses. This is in contradiction to Visser et~al. (2001) who calculate the mass of L675 to be
$0.2$\,M$_{\odot}$, making it the lowest mass object in our
sample.

\begin{figure}
\centerline{\includegraphics[height=6.5cm,width=8.cm,angle=0]{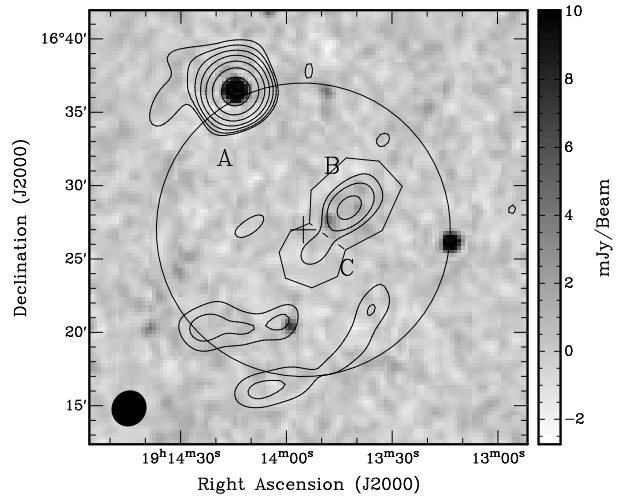}}
\caption{L709: CLEANed AMI combined channel data at 16\,GHz are shown as contours with the
  levels as in Fig.~\ref{fig:l675}. NVSS data at 1.4\,GHz are shown
  in greyscale and are saturated at 10\,mJy peak to emphasise the
  structure in the map. The AMI pointing centre is indicated by a cross and
  the FWHM of the primary beam as a solid circle. The synthesized beam
  of AMI towards L709, $2\farcm5 \times 2\farcm4$, is shown in the bottom
  left corner as a filled ellipse. This map is not primary beam
  corrected. Example apertures are
  shown to indicate the division between sources ``B'' and ``C'' for
  flux extraction.\label{fig:l709}
}
\end{figure}

\subsection{L709:} 
\begin{figure}
\centerline{\includegraphics[height=8.cm,angle=-90]{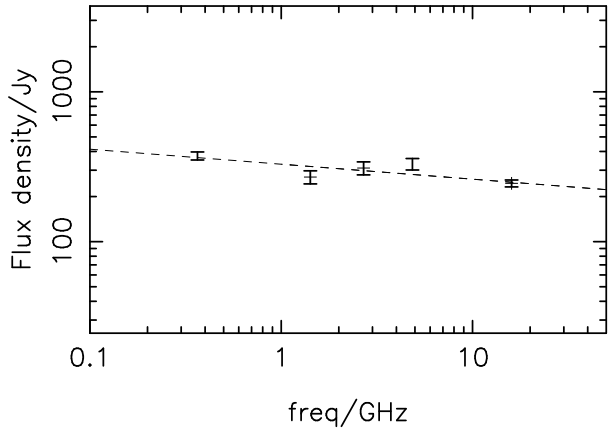}}
\caption{Radio spectrum of the L709 point source ``A''. These data
  points are described in the text.\label{fig:l709a}}
\end{figure}

There is no microwave source coincident with the
  pointing centre towards L709, see Fig~\ref{fig:l709}. The
field is dominated by a bright radio source to the north-east of the
pointing centre, denoted ``A'' in Fig.~\ref{fig:l709}. Source ``A'' lies slightly
outside the FWHM of the AMI primary beam and has a primary beam
corrected continuum flux of $S_{16} =
245\pm13$\,mJy. This is consistent with fluxes from the literature of
$S_{0.365} = 374\pm24$\,mJy (Texas), $S_{1.42}=270\pm27$\,mJy (Effelsberg
21cm), $S_{2.695}=310\pm31$\,mJy (Effelsberg 11cm) and
$S_{4.85}=330\pm29$\,mJy (GB6) indicating a source with spectral index
$\alpha = 0.07\pm0.02$, see Fig.~\ref{fig:l709a}. The NVSS flux
density for this object is slightly lower, $S_{\rm{NVSS},1.4} =
218\pm6.6$\,mJy suggesting that the source is extended. A second, and fainter, object in the field is
just north-west of the pointing centre, Source ``B''. This object has a flux density of
$S_{16} = 6.0\pm0.5$\,mJy in the AMI map and is most probably a
combination of three unresolved point sources, which may be found in
the NVSS catalogue, all falling within an AMI synthesized beam at this
position. These sources have a
combined flux density of $S_{1.4} = 11.0\pm0.5$\,mJy, indicating a
decrease in flux at 16\,GHz. The slight extension
to the south of this object, ``C'', is not coincident
with any NVSS point sources and may be associated with L709. The
object is present at the 
5$\sigma$ level in the combined channel map with a flux of $S_{16} =
1.61\pm0.31$\,mJy. In the consitituent channel maps the source is
present with flux densities varying between 1 and 4\,$\sigma$. At this
level of significance it is difficult to fit a reliable spectral
index, however the flux density appears to be steeply falling with
increasing frequency. As the source appears point-like this would
suggest that Source ``C'' is not associated with L709 but is instead a
faint steep spectrum extragalactic point source. 

\begin{figure}
\centerline{\includegraphics[height=6.5cm,width=8.cm,angle=0]{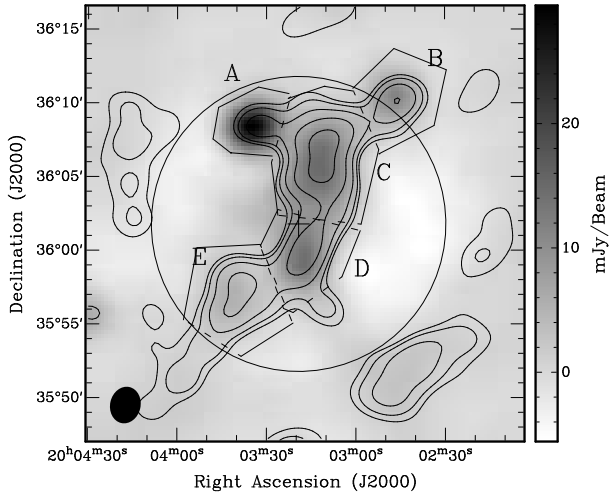}}
\caption{L860: CLEANed combined channel AMI data at 16\,GHz are shown as contours with the
  levels as in Fig.~\ref{fig:l675}. {\it{uv}} sampled CGPS data at 1.4\,GHz are shown
  in greyscale. The AMI pointing centre is indicated by a cross and
  the FWHM of the primary beam as a solid circle. The synthesized beam
  of AMI towards L860, $2\farcm6 \times 2\farcm1$, is shown in the bottom
  left corner as a filled ellipse. This map is not primary beam
  corrected. Example apertures for
  flux extraction are shown to illustrate the division between the
  regions denoted A--E.\label{fig:l860}
}
\end{figure}

\subsection{L860:}\label{sec:l860} At 16\,GHz we see a ridge of
  emission towards L860, see Fig.~\ref{fig:l860}. We investigate its spectral
properties using analysis Case (1). Although it possesses no obvious counterpart in the
unsampled CGPS dataset, a \emph{uv} matched image shows the same
structure at 1.4\,GHz. To the north of the field two point-like radio 
sources may be found (A \& B), whilst the ridge of extended emission
that runs north--south across the pointing centre may be divided into
three distinct sub-regions of emission (C, D \& E). The morphology of
these sub-regions is not well described by a Gaussian model.
Although both C \& D might be considered to be associated with L860 the derived flux densities, see Table~\ref{tab:list}, and their
spectral indices 
indicate that there is no excess emission present at microwave
frequencies for these sources.

\begin{table*}
\caption{Flux densities for compact sources found within the AMI fields
  which have counterparts at 1.4\,GHz.\label{tab:list}
}\begin{tabular}{lccccc}
\hline
Source & RA & Dec & $S_{1.4}$ & $S_{16}$ &
$\alpha_{1.4}^{16}$ \\
&(2000.0)&(2000.0)&(mJy)&(mJy)&\\
\hline
\hline
L709&&&&&\\
A & 19 14 14.8 & 16 36 43 & 270$\pm$27 & 245$\pm$13 & 0.07$\pm$0.02\\
L860&&&&&\\
A & 20 03 34.2 & 36 08 30 & 43.4$\pm$5.0 & 5.2$\pm$1.1 & 0.88$\pm$0.30\\
B & 20 02 44.6 & 36 10 30 & 23.2$\pm$2.6 & 14.1$\pm$2.3 & 0.21$\pm$0.11\\
C & 20 03 10.7 & 36 07 15 & 36.6$\pm$1.2 & 35.6$\pm$2.3 & 0.01$\pm$0.01 \\
D & 20 03 16.8 & 35 59 60 & 25.0$\pm$4.1 & 19.6$\pm$1.5 & 0.10$\pm$0.07 \\
E & 20 03 41.5 & 35 55 60 & 17.9$\pm$2.4 & 17.4$\pm$2.8 & 0.01$\pm$0.03 \\
L917&&&&&\\
A & 20 39 31.5 & 44 02 11 & 64.9$\pm$6.3 & 66.2$\pm$11.7 & $-$0.01$\pm$0.03\\
B & 20 39 32.8 & 44 09 41 & 2.8$\pm$0.7 & 15.0$\pm$1.6 & $-$0.69$\pm$0.21\\
C & 20 39 43.9 & 44 14 11 & 15.6$\pm$3.2 & 17.9$\pm$2.6 & $-$0.06$\pm$0.07 \\
L944&&&&&\\
A & 21 18 01.4 & 43 16 53 & 11.0$\pm$0.5 & 1.3$\pm$0.3 & 0.89$\pm$0.79\\
B & 21 17 23.0 & 43 11 08 & 17.8$\pm$0.7 & 5.7$\pm$0.9 & 0.47$\pm$0.12\\
C & 21 18 34.3 & 43 13 22 & 35.1$\pm$1.1 & 9.9$\pm$0.9 & 0.52$\pm$0.10 \\
D & 21 16 45.9 & 43 13 37 & 57.6$\pm$1.8 & 7.7$\pm$0.9 & 0.83$\pm$0.17 \\
E & 21 16 47.1 & 43 23 52 & 25.3$\pm$0.9 & 7.4$\pm$0.9 & 0.50$\pm$0.11 \\
L951&&&&&\\
A & 21 20 15.8 & 43 37 15 & 190.2$\pm$7.5 & 23.5$\pm$4.4 & 0.89$\pm$0.79\\
B & 21 19 48.1 & 43 36 45 & 60.7$\pm$2.2 & 9.5$\pm$2.2 & 0.47$\pm$0.12\\
C & 21 19 23.3 & 43 39 30 & 15.7$\pm$1.1 & 2.8$\pm$0.9 & 0.52$\pm$0.10 \\
D & 21 20 30.6 & 43 19 56 & 4.7$\pm$0.5 & 23.9$\pm$5.2 & 0.83$\pm$0.17 \\
L953&&&&&\\
A & 21 20 30.6 & 43 19 56 & 4.7$\pm$0.5 & 20.3$\pm$4.8 & 0.60$\pm$0.15 \\
L1014&&&&&\\
A & 21 23 49.2 & 49 59 11 & 44.5$\pm$1.1 & 8.5$\pm$1.4 & 0.47$\pm$0.12\\
B & 21 24 01.7 & 49 56 11 & 67.7$\pm$2.4 & 3.9$\pm$1.4 & 0.52$\pm$0.10 \\
C & 21 23 19.7 & 49 59 11 & 194.3$\pm$6.9 & 25.5$\pm$2.6 & 0.89$\pm$0.79\\
D & 21 24 39.0 & 50 00 26 & 18.4$\pm$0.7 & 4.1$\pm$1.4 & 0.83$\pm$0.17 \\
E & 21 24 53.0 & 49 57 11 & 5.9$\pm$0.5 & 4.8$\pm$1.5 & 0.50$\pm$0.11 \\
F & 21 23 04.1 & 50 02 26 & 194.3$\pm$6.9 & 25.3$\pm$1.5 & 0.50$\pm$0.11 \\
L1021&&&&&\\
A & 21 21 45.1 & 50 54 07 & 7.6$\pm$0.5 & 4.1$\pm$1.3 & 0.89$\pm$0.79\\
L1103&&&&&\\
A & 21 43 06.5 & 56 40 52 & 140.0$\pm$3.9 & 13.4$\pm$4.4 & 0.97$\pm$0.65\\
B & 21 41 58.4 & 56 36 30 & 17.6$\pm$5.0 & 4.0$\pm$2.1 & 0.61$\pm$0.46\\
C & 21 41 36.5 & 56 46 54 & 29.5$\pm$5.9 & 22.1$\pm$4.8 & 0.12$\pm$0.14 \\
D & 21 42 13.2 & 56 48 28 & 24.3$\pm$2.1 & 4.3$\pm$2.3 & 0.72$\pm$0.55 \\
E & 21 41 34.7 & 56 53 22 & 21.7$\pm$2.3 & 4.3$\pm$2.4 & 0.67$\pm$0.54 \\
F & 21 41 26.5 & 56 53 48 & 20.1$\pm$3.3 & 15.8$\pm$7.7 & 0.10$\pm$0.20 \\
L1111&&&&&\\
B & 21 39 55.0 & 57 48 59 & 12.2$\pm$1.3 & 3.0$\pm$0.6 & 0.58$\pm$0.14\\
L1185&&&&&\\
A & 22 29 26.5 & 59 07 38 & 66.1$\pm$1.3 & 2.6$\pm$1.6 & 1.33$\pm$1.11\\
B & 22 28 12.2 & 59 13 37 & 55.1$\pm$1.3 & 63.5$\pm$5.9 & $-$0.06$\pm$0.04\\
L1246&&&&&\\
A & 23 24 46.0 & 63 29 59 & 9.0$\pm$0.9 & 5.6$\pm$1.1 & 0.20$\pm$0.08\\
B & 23 24 21.4 & 63 28 14 & 41.5$\pm$1.3 & 2.4$\pm$0.5 & 1.17$\pm$0.31\\
\hline
\end{tabular}
\begin{minipage}{16cm}
{}
\end{minipage}
\end{table*}

\begin{figure*}
\centerline{\includegraphics[height=6.5cm,width=8.cm,angle=0]{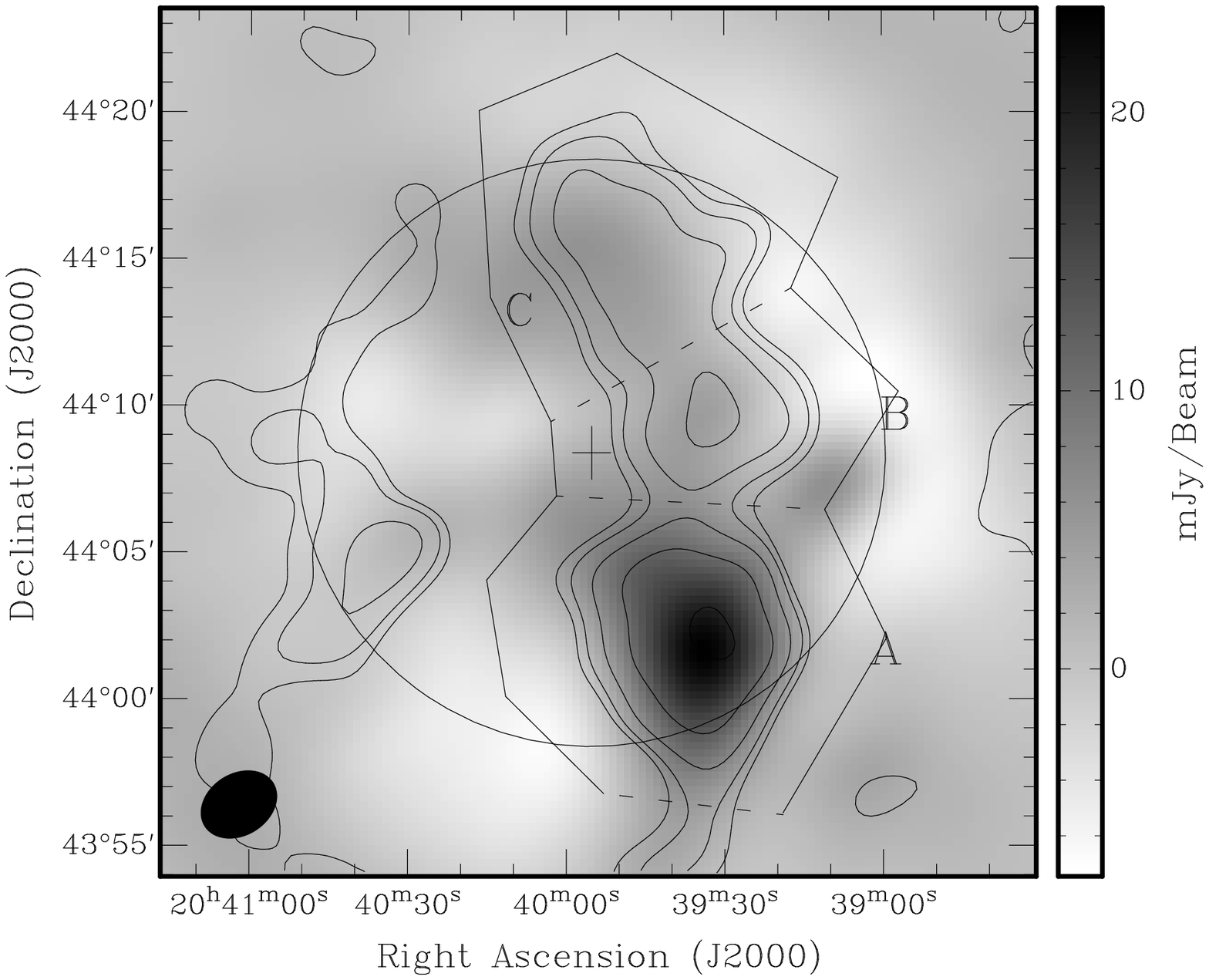}\qquad{\includegraphics[height=6.5cm,width=8.cm,angle=0]{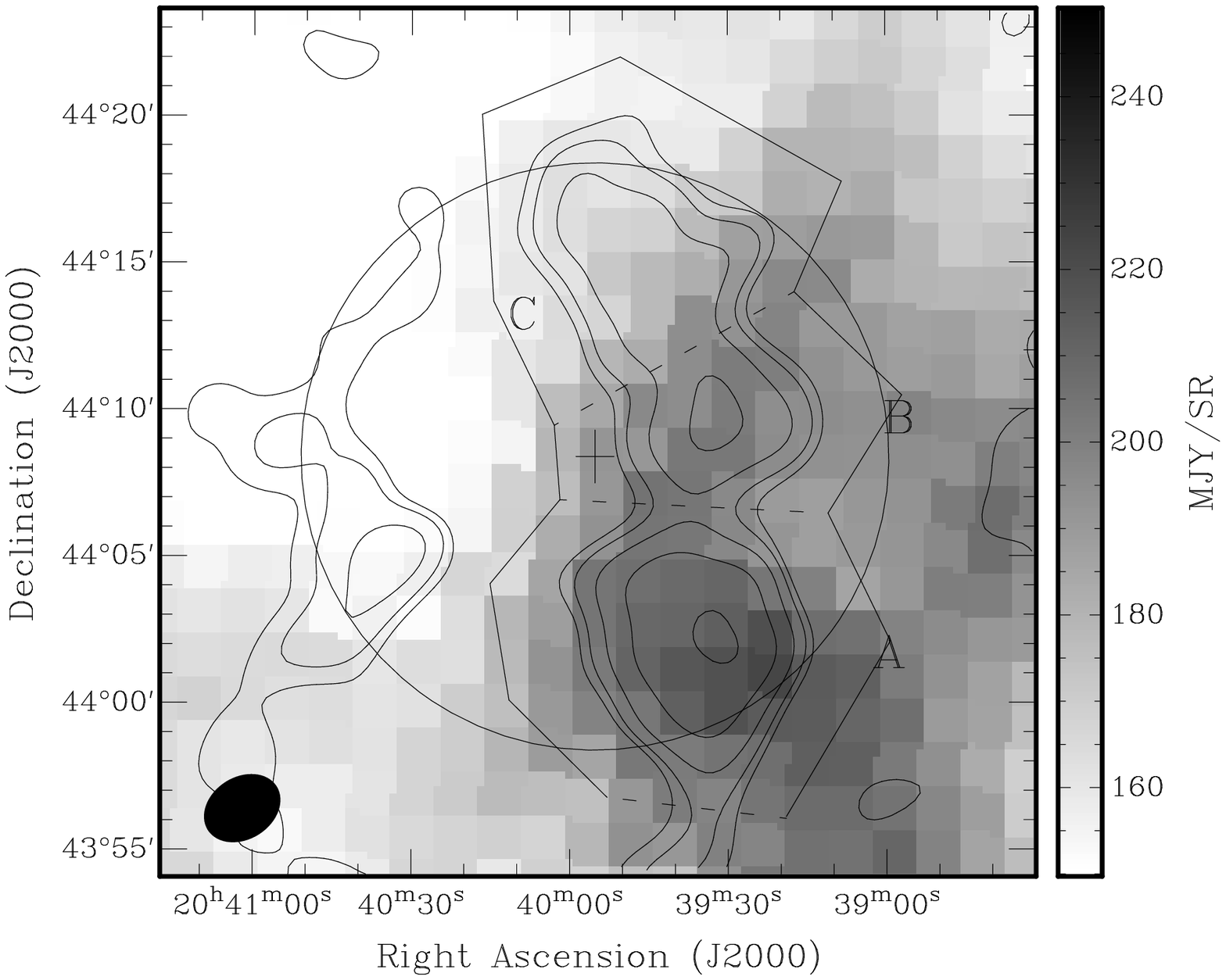}}}
\caption{L917: (a) Combined channel AMI data at 16\,GHz are shown as contours with the
  levels as in Fig.~\ref{fig:l675}. {\it{uv}} matched CGPS data at 1.4\,GHz are shown
  in greyscale. (b) AMI data at 16\,GHz are shown as contours with the
  levels as in Fig.~\ref{fig:l675}. 100\,$\mu$m IRAS data are shown
  in greyscale. The AMI pointing centre is indicated by a cross and
  the FWHM of the primary beam as a solid circle. These maps are not
  primary beam corrected. The AMI synthesized beam, $2\farcm8 \times 2\farcm1$, is shown as a
  filled ellipse in the bottom left corner. Example flux
  extraction apertures are shown to illustrate the divisions between
  the 
  regions denoted A--C.\label{fig:l917i12}
}
\end{figure*}

\subsection{L917:} 
\begin{figure}
\centerline{\includegraphics[height=8.cm,angle=-90]{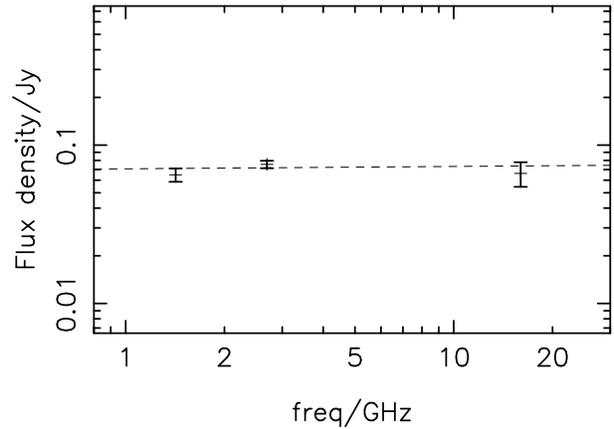}}
\vspace{0.3cm}

\centerline{\includegraphics[height=8.cm,angle=-90]{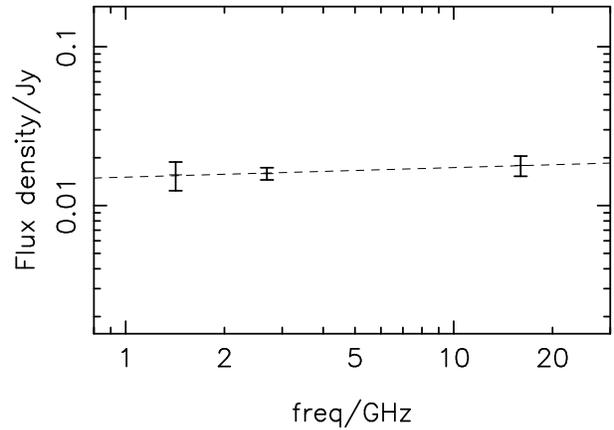}}
\caption{Above: Radio spectrum of L917 region ``A''. Below: Radio spectrum of L917 region ``C''. Data points
  are {\it{uv}} matched flux densities at 1.4, 2.7 and 16\,GHz from
  the CGPS, Effelsberg 100\,m telescope and AMI, respectively. See
  text for details.\label{fig:l917spec}
 }
\end{figure}

No radio emission can be seen directly towards L917
at 16\,GHz,
although a ridge of emission runs north--south slightly to the west of
the pointing centre, see Fig.~\ref{fig:l917i12}. We investigate the
spectral properties of this ridge primarily using analysis Case (1). This ridge has three separate peaks, which are
evident in both the 1.4\,GHz {\it{uv}} sampled data and the AMI data
at 16\,GHz. We fit for the flux density of each peak separately using
the flux extraction method described in Section~\ref{sec:l675}. These
peaks (A, B \& C) all appear to have slightly more flux at
16\,GHz than at 1.4\,GHz, see Table~\ref{tab:list}. Using additional
data at 2.7\,GHz from the Effelsberg telescope {\it{uv}} sampled under
Case (2) we can fill in more of the flux spectrum. The peaks A
and C show a spectrum consistent with a region of optically thin
free--free emission, see Fig.~\ref{fig:l917spec}. In ``B'', the closest peak to the pointing
centre we see a large excess at 16\,GHz relative to 1.4\,GHz, although
it is not clear if this 
excess is caused by anomalous emission. At 2.7\,GHz the emission has a
largely different morphology making it difficult to constrain a
spectrum using the three frequencies. A second factor which makes
emission from spinning dust unlikely in this case is the spectral
index across the AMI channels, $\alpha_{14.2}^{17.9} =
0.54\pm0.36$. With a falling spectrum this region is not a good
candidate for a spinning dust emission, even considering a correction
for flux loss. It can be seen from the IRAS
data at all four frequencies that there is a ridge of dust which is
spatially associated with regions A and B, see
Fig.~\ref{fig:l917i12}. However, in spite of this there is no formal
IRAS association with L917. The change in morphology between the radio 
frequencies could be accounted for by this region being composed of
a number of small {\sc{Hii}} regions with different turn-over
frequencies into the optically thin regime.

\begin{figure}
\centerline{\includegraphics[height=6.5cm,width=8.cm,angle=0]{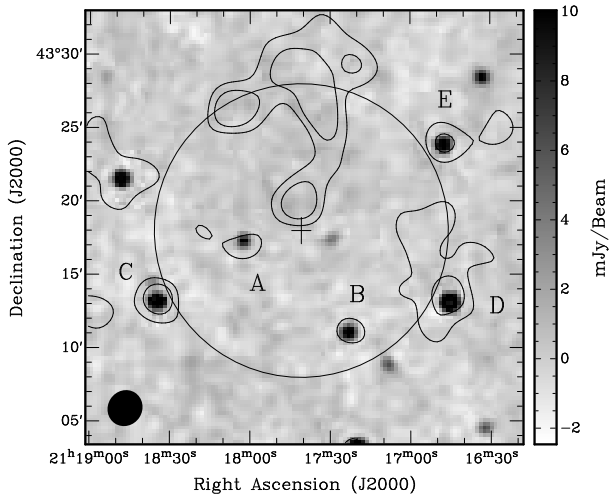}}
\caption{L944: CLEANed combined channel AMI data at 16\,GHz are shown as contours with the
  levels as in Fig.~\ref{fig:l675}. NVSS data at 1.4\,GHz are shown
  in greyscale and are saturated at 10\,mJy peak. The AMI pointing centre is indicated by a cross and
  the FWHM of the primary beam as a solid circle. This map is not
  primary beam corrected. The synthesized beam
  of AMI towards L944, $2\farcm5 \times 2\farcm3$, is shown in the bottom
  left corner as a filled ellipse.\label{fig:l944}}
\end{figure}

\subsection{L944:} The field around L944 contains a number of compact
 radio sources which can be seen in both the AMI data and the
corresponding NVSS map, see Fig.~\ref{fig:l944}. Five sources near
to the pointing centre all have non-thermal spectral
indices between 1.4 and 16\,GHz consistent with those extrapolated
from the WENSS survey at 327\,MHz.
One source is found in the
AMI map just north of the pointing centre which possesses no
counterpart in the NVSS data. This source is coincident with the
outflow from L944 which is extended towards the North and has a flux
density of $S_{16} = 3.5\pm0.3$\,mJy in the combined channel AMI map. 

The spectrum of the this source appears relatively flat across the AMI
band, $\alpha_{14.2}^{17.9} = -0.07\pm1.04$, and may be a thermal source which remains optically thick to
high frequencies turning over below 16\,GHz. It has no counterpart in either the NVSS data at
1.4\,GHz, {\it{uv}} sampled Effelsberg data at 2.7\,GHz or the GB6
data at 4.85\,GHz. It is not possible to constrain
the spectral index of the possible optically thick component, as the
source flux at 16\,GHz already lies below the noise level for both the
Effelsberg data and the GB6 survey. 

The short frequency lever arm of the AMI data alone makes an estimate
of the spectral index from 14.2 to 17.9\,GHz imprecise. However,
including the constraint provided by the lack of any flux at 1.4\,GHz
in the NVSS data constrains the spectral index to be
$\alpha_{1.4}^{17.9} \leq -1.2$. Such a spectral index could 
be consistent with the optically thick regime of an
ultracompact/hypercompact {\sc{Hii}} region, the spectral indices of which are normally slightly shallower than $\alpha = -2$
(Franco et~al. 2000).

Two obvious possibilities to explain these observations might also be cm-wave emission from a
protostar within L944, or alternatively emission from a protoplanetary
disc which may also extend into the cm regime. The second
of these seems unlikely: protoplanetary discs are associated with
later T-Tauri stars and are also extremely small in angular
diameter. For the flux density
measured at 16\,GHz to be accounted for by the tail of the dust SED we
would require $\beta = 0.1$, consistent with the SED of
a protoplanetary disc. However, we find that the flux densities from
SCUBA and IRAS are best fitted by $\beta=0.8$ in agreement with
Froebrich (2005). Assuming a source point-like to the IRAS beam, this
would predict a flux density from thermal 
dust of only 0.55\,mJy at 16\,GHz.

The possibility of cm-wave emission from a protostar is not
unlikely. L944 is known to be a Class 0 
source, having extended and centrally peaked submillimetre continuum
emission, which indicates a spheroidal envelope; a high ratio of
submillimetre to bolometric luminosity (Froebrich 2005) and bipolar
molecular outflows (Visser et~al. 2002). Given these conditions it
should also have an internal heating source, which would give rise to
compact cm-wave emission. Previous observations of candidate
protostars in the cm regime (e.g. Andre et~al. 1993; Stamatellos
et~al. 2007) found fluxes which would not be inconsistent with that
measured by AMI towards this source, however we note that the observed AMI
flux densities would be higher than the average flux density.

\begin{figure*}

\centerline{\includegraphics[height=6.5cm,width=8.cm,angle=0]{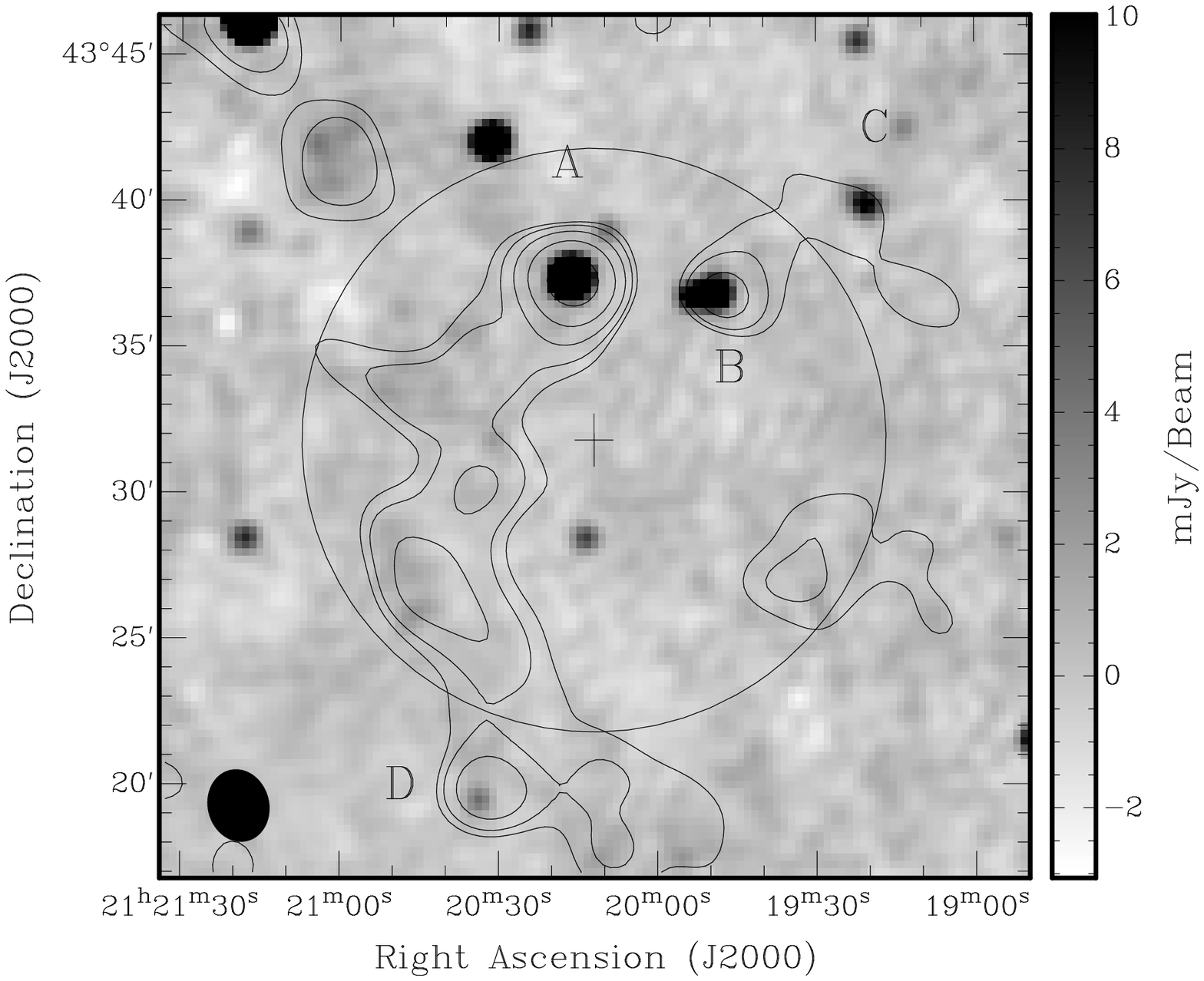}\qquad\includegraphics[height=6.5cm,width=8.cm,angle=0]{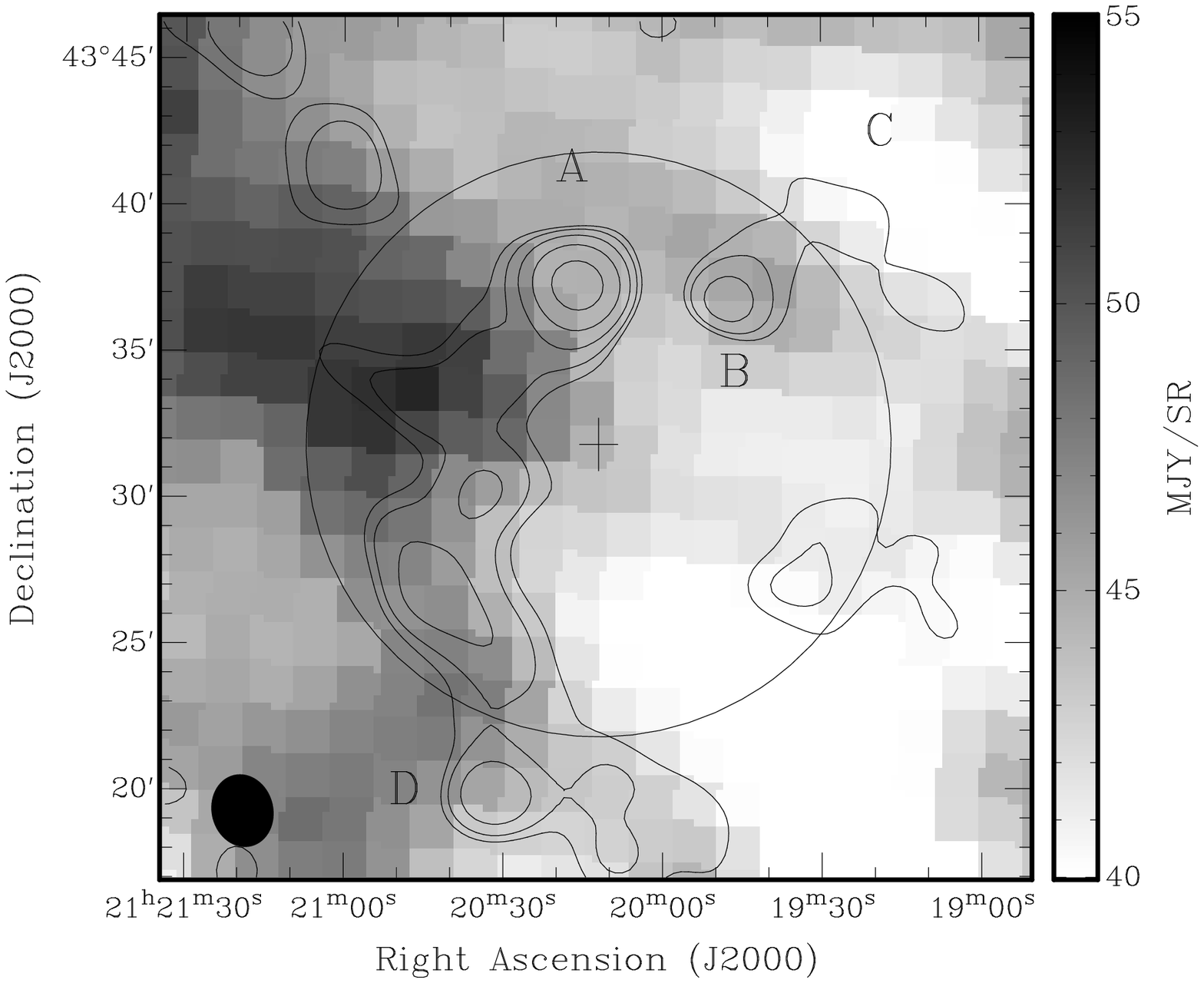}} 
\centerline{(a) \hspace{7cm} (b) }
\caption{L951: (a) CLEANed combined channel AMI data at 16\,GHz are shown as contours with the
  levels as in Fig.~\ref{fig:l675}. CLEANed {\it{uv}} matched CGPS data at 1.4\,GHz are shown
  in greyscale. (b) AMI data at 16\,GHz are shown as contours with the
  levels as in Fig.~\ref{fig:l675}. 100\,$\mu$m IRAS data are shown
  in greyscale. The AMI pointing centre is indicated by a cross and
  the FWHM of the primary beam as a solid circle. The AMI synthesized
  beam, $2\farcm5 \times 2\farcm2$, is shown as a filled ellipse in
  the bottom left corner.\label{fig:l951}
}
\end{figure*}

\begin{figure}
\centerline{\includegraphics[height=6.5cm,width=8.cm,angle=0]{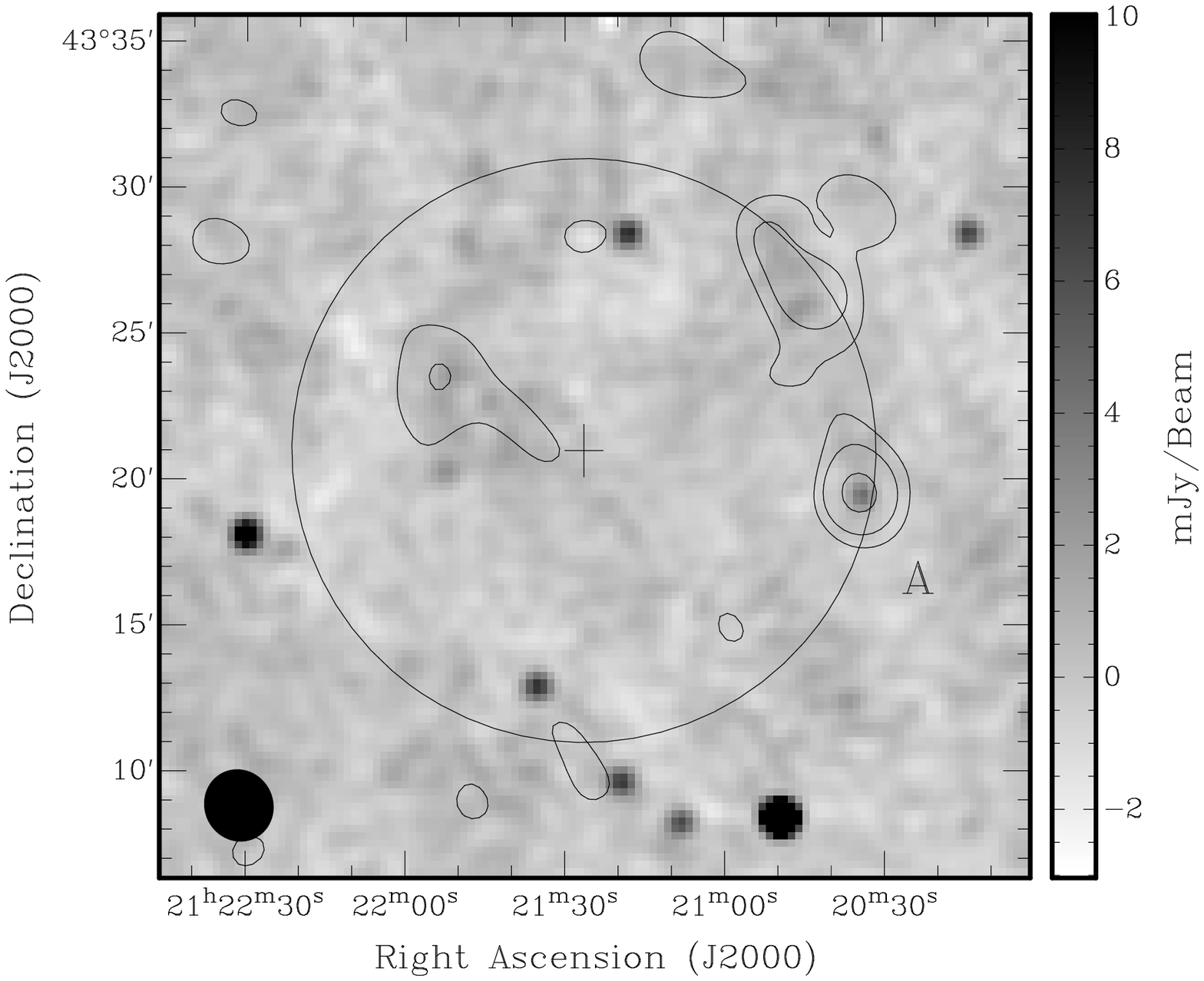}}
\caption{L953: CLEANed combined channel AMI data at 16\,GHz are shown as contours with the
  levels as in Fig.~\ref{fig:l675}. NVSS data at 1.4\,GHz are shown
  in greyscale and are saturated at 10\,mJy peak. The AMI pointing centre is indicated by a cross and
  the FWHM of the primary beam as a solid circle. The AMI synthesized
  beam, $2\farcm5 \times 2\farcm3$, is shown as a filled ellipse in
  the bottom left corner.\label{fig:l953}
}
\end{figure}

\subsection{L951 \& L953:} 

There is no radio emission evident towards either L951 or L953, see
Fig.~\ref{fig:l951} and Fig.~\ref{fig:l953}. The
two fields overlap by a small amount -- source ``D'' in the L951 field
is also source ``A'' in the L953 field. The flux of this source agrees
well between the observations and although it appears to have a rising
spectrum from 1.4 to 16\,GHz it has turned over by 4.85\,GHz and may
be found in the GB6 catalogue with a flux of $S_{4.85} = 36\pm5$\,mJy,
giving a spectral index of $\alpha_{4.85}^{16} = 0.41\pm0.14$. The
ridge of extended emission which runs roughly north--south along the
eastern edge of the L951 field and the western edge of the L953 field
may be associated with an enhancement of the local infrared emission
and may be seen in IRAS data towards the same region, see
Fig.~\ref{fig:l951}\,(b).

\begin{figure}
\centerline{\includegraphics[height=6.5cm,width=8.cm,angle=0]{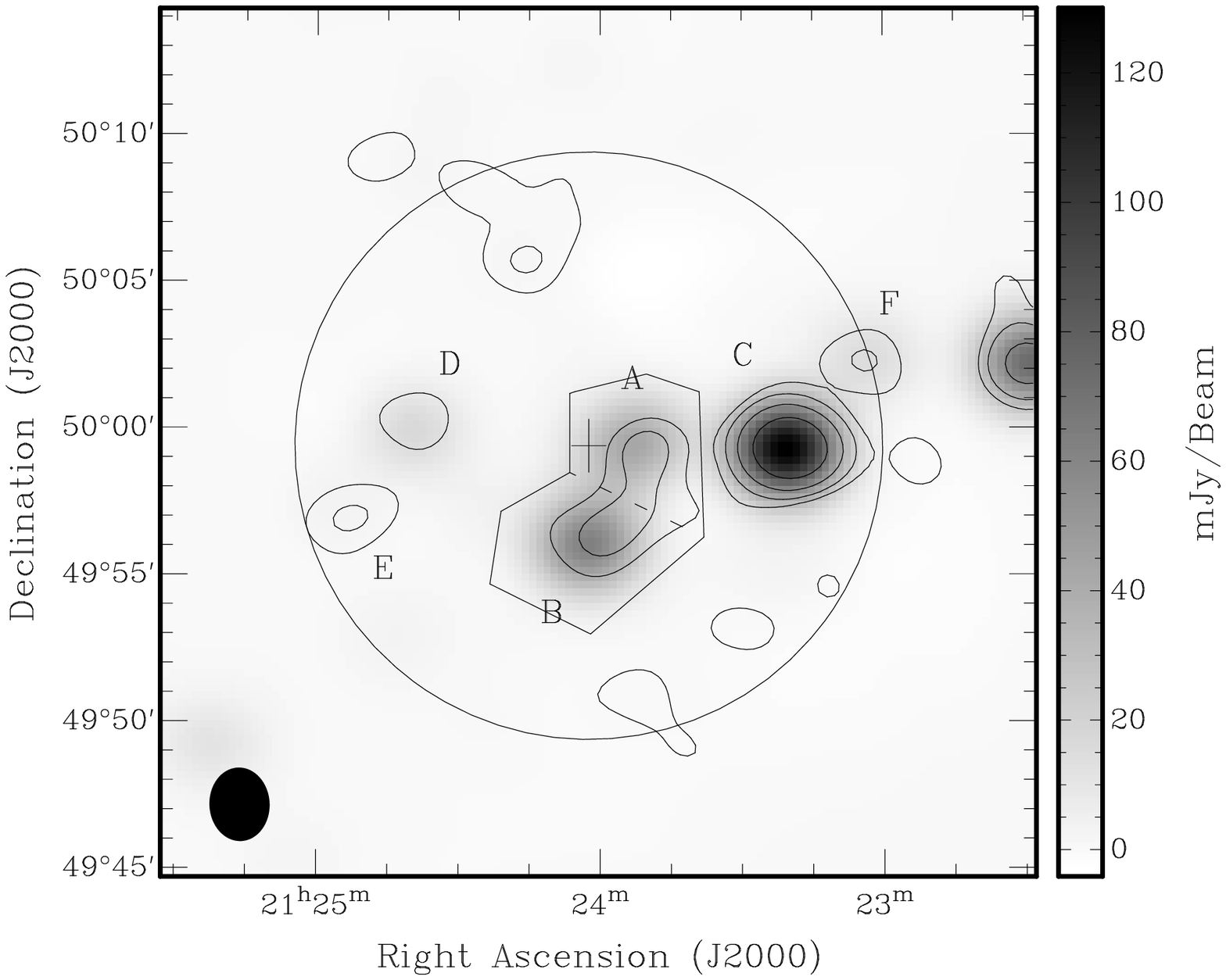}}\caption{L1014:
  CLEANed combined channel AMI data at 16\,GHz are shown as contours with the
  levels as in Fig.~\ref{fig:l675}. CLEANed {\it{uv}} matched data from the
  CGPS at 1.4\,GHz are shown
  in greyscale. The AMI pointing centre is indicated by a cross and
  the FWHM of the primary beam as a solid circle. The AMI synthesized
  beam, $2\farcm5 \times 2\farcm1$, is shown as a filled ellipse in
  the bottom left corner. Example flux
  extraction apertures are shown to illustrate the division between
  sources ``A'' and ``B''.\label{fig:l1014}
}

\end{figure}

\subsection{L1014:} The L1014 field contains a number of
compact sources none of which is completely coincident with the SCUBA
position of the dark
cloud, see Fig.~\ref{fig:l1014}. The two closest sources ``A'' and
``B'' both have falling spectral indices and are therefore not
candidates for anomalous emission. Like
L944, L1014 is a Class 0 source (Young et~al 2004; Bourke
et~al. 2005). Radio cm-wave observations with the VLA have identified
a candidate for the protostar at RA 21h24m07.53s, $\delta$ =
+49$^{\circ}$59'08.9'' (J2000.0) (Shirley et~al. 2007). Their
measurements would imply a flux density of $S_{16} = 145\mu$Jy at
16\,GHz, well below the thermal noise in this observation.

\begin{figure}
\centerline{\includegraphics[height=6.5cm,width=8.cm,angle=0]{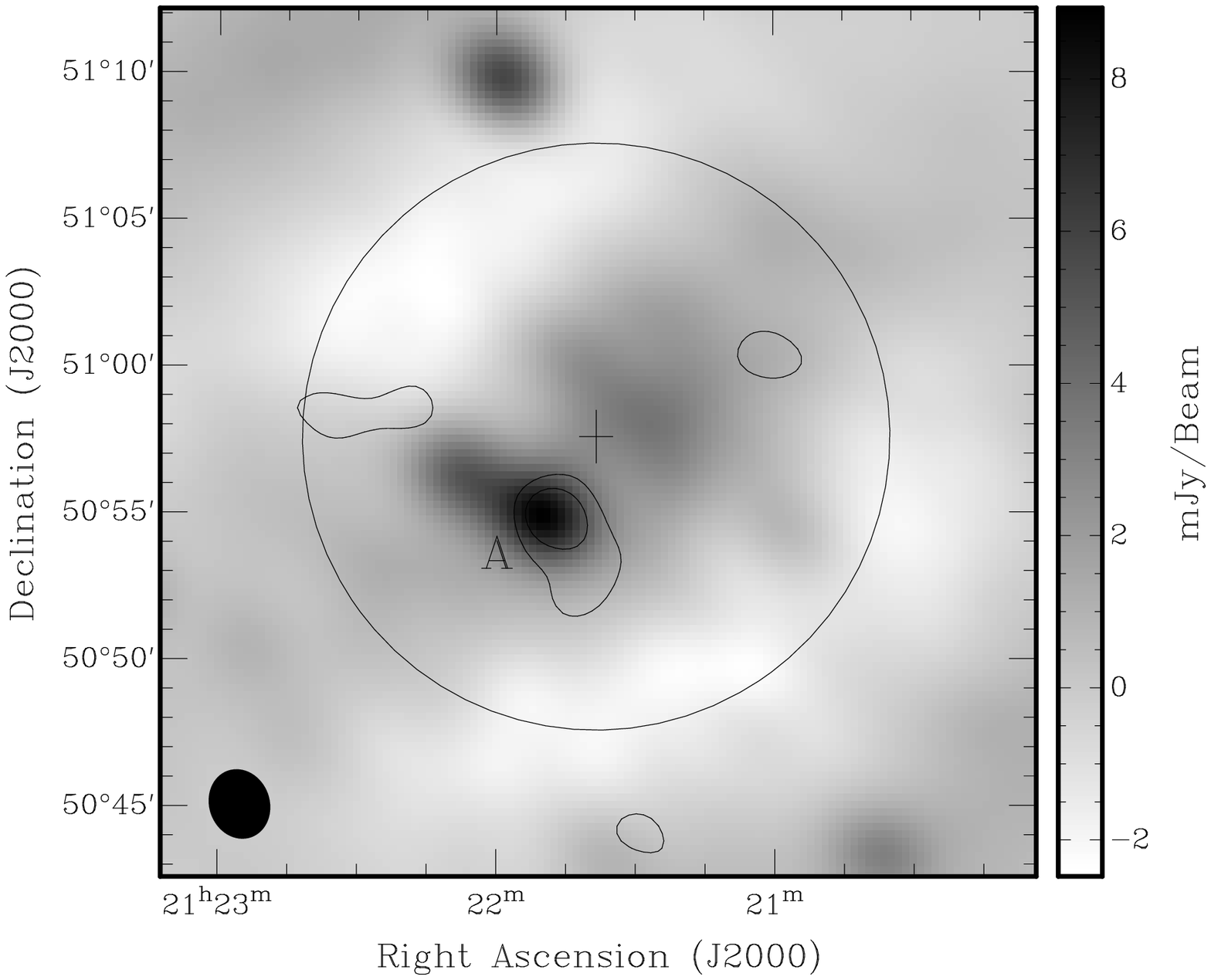}}\caption{L1021:
  CLEANed combined channel AMI data at 16\,GHz are shown as contours with the
  levels as in Fig.~\ref{fig:l675}. CLEANed {\it{uv}} matched data from the
  CGPS at 1.4\,GHz are shown
  in greyscale. The AMI pointing centre is indicated by a cross and
  the FWHM of the primary beam as a solid circle. The AMI synthesized
  beam, $2\farcm4 \times 2\farcm1$, is shown as a filled ellipse in
  the bottom left corner.\label{fig:l1021}
}

\end{figure}

\subsection{L1021:} This relatively empty field contains only one
radio source which can also be identified in the lower frequency
data, see Fig.~\ref{fig:l1021}. Its falling spectral index suggests
that it is a non-thermal extragalactic point source.

\begin{figure}
\centerline{\includegraphics[height=6.5cm,width=8.cm,angle=0]{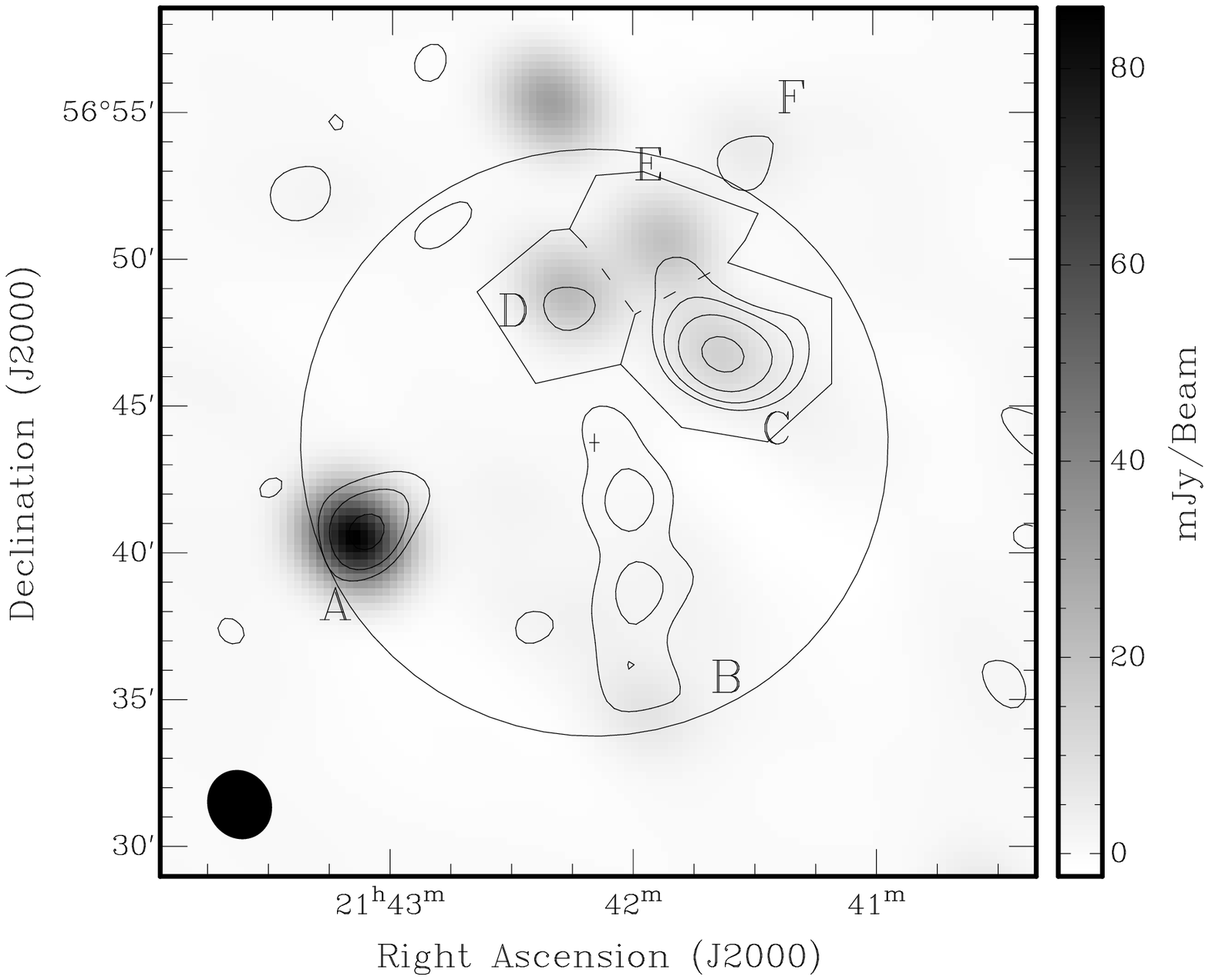}}
\caption{L1103: CLEANed combined channel AMI data at 16\,GHz are shown as contours with the
  levels as in Fig.~\ref{fig:l675}. CLEANed {\it{uv}} matched data from the
  CGPS at 1.4\,GHz are shown
  in greyscale. The AMI pointing centre is indicated by a cross and
  the FWHM of the primary beam as a solid circle. The AMI synthesized
  beam, $2\farcm4 \times 2\farcm1$, is shown as a filled ellipse in
  the bottom left corner. Example flux
  extraction apertures are shown to illustrate the division between
  those sources labelled C--E. \label{fig:l1103}
}
\end{figure}

\begin{figure}
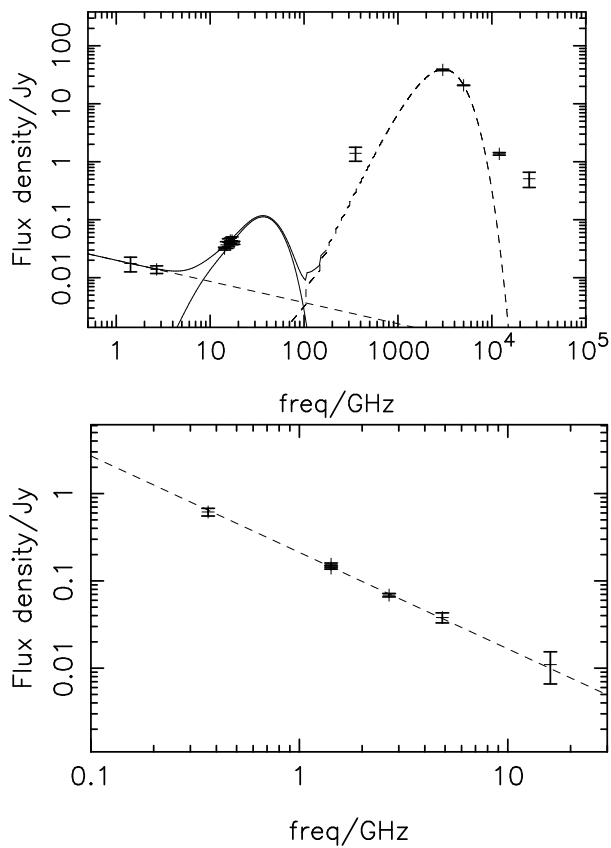


\centerline{\includegraphics[height=8.cm,angle=-90]{./L1103_spec.ps}}
\centerline{\includegraphics[height=8.cm,angle=-90]{./L1103a_spec.ps}}
\caption{Top: SED of L1103.  Data points are shown at 1.4 and 2.7\,GHz from the
  CGPS and Effelsberg 100\,m. These
  data are {\it{uv}} matched to the data at 13.8 to 17.6\,GHz
  from AMI which are also shown. As are points at 850\,$\mu$m (353\,GHz) from
  Visser et~al. (2001; 2003); and at 100, 60, 25 and 12\,$\mu$m from
  Clemens, Yuo \& Heyer (1991). A modified greybody spectrum with
  $\beta = 1.5$ is fitted to the 850, 100 and 60\,$\mu$m data as
  described in the text and a MC spinning dust model from DL98 is
  fitted to the AMI data points, see text for details. The combined SED is shown as a solid
  line. Bottom: SED of J214307+564033. Data points are shown at
  327\,MHz (WENSS, Rengelink et~al. 1997), 1.4\,GHz (NVSS, Condon et~al. 1998;
  {\it{uv}} matched CGPS, this work), 2.7\,GHz ({\it{uv}} matched
  Effelsberg 100\,m, this work), 4.85\,GHz (GB6, Gregory et~al. 1996) and
  16\,GHz (AMI, this work). A best fitting spectral index of $\alpha
  =1.1\pm0.26$ is shown as a dashed line.\label{fig:l1103spec}
  }
\end{figure}

\subsection{L1103:} This field contains a number of small diameter
sources, the brightest of which, ``A'', we identify as J214307+564033. Sampling the corresponding CGPS data towards L1103 reveals low
frequency counterparts to many of these, see Fig~\ref{fig:l1103}. However, the object found closest to
the pointing centre of this field has no lower
frequency counterpart. The maximum flux towards the pointing centre
from the {\it{uv}} sampled CGPS data at 1.4\,GHz is 1\,mJy\,beam$^{-1}$, compared
to a peak flux density of 4\,mJy\,beam$^{-1}$ at 16\,GHz. Moreover, an extended
structure is obvious in the AMI data running south from the pointing
centre which is not evident at 1.4\,GHz. The total integrated flux of
this object from the combined AMI channels is $40.1\pm3.1$\,mJy. The ridge is present at all AMI frequencies
and has a rising spectral index of $\alpha = -1.6\pm0.3$. However the
ridge has 
dimensions of approximately $10'\times 3'$, and at such sizes we would
expect a significant amount of flux loss across the AMI band. This
flux loss would have the effect of steepening a falling spectrum, or
flattening a rising spectrum. Since
the morphology in the 1.4\,GHz data is very different we cannot use
this as a model to calculate absolute flux loss. Instead we model the
ridge as a multivariate Gaussian and calculate a flux loss correction
for channels 4--8, relative to channel 3. This will give us the true
slope across the AMI band, if not the total power normalization. With
these corrections we find a corrected spectral index of $\alpha =
-2.2\pm0.2$. A 
further correction we might make is to remove the flux of the point like
object ``B'' from the flux density of the ridge as a whole. However,
it is unclear from a comparison of the AMI and {\it{uv}} sampled CGPS
data how large the contribution of B to the ridge flux density may
be. Although the ridge certainly extends to meet B, there is no clear
peak at the position of the 1.4\,GHz source. Consequently, since the
difference in flux density across the AMI band will be small compared
to the necessary increase in the error budget we chose not to subtract
the flux of this object from that of the ridge as a whole. Instead we
consider object ``B'' as part of the ridge and simply find the
integrated flux density of the entire ridge. At 1.4\,GHz and 2.7\,GHz
there is no apparent elongation towards the pointing centre and so the
flux density is heavily dominated by the southern end of the
ridge. Using {\it{uv}} matched data at 1.4 and 2.7\,GHz we find a
poorly constrained spectral index of $\alpha_{1.4}^{2.7} = 0.37\pm0.30$. Again we urge
caution when considering the 2.7\,GHz flux density as it is only
marginally above the noise level of the original Effelsberg data. 

Although, like L675, L1103 has no formal IRAS association it was also
observed using deeper IRAS photometry by Clemens, Yun \& Heyer (1991)
and we use their measurements to constrain the FIR SED. They found the size of L1103 to be $\Omega_S = 77$\,arcmin$^2$, more
consistent with the area seen by AMI than with SCUBA. Again as in the
case of L675, this is perhaps not surprising given the SCUBA 120$''$
chop. Note that the greybody spectrum which best
fits the IRAS data points at 60 and 100\,$\mu$m significantly
underestimates the SCUBA flux at 850$\mu$m when the differing angular
size of the observed source is taken into account. We find a best
fitting SED with $\beta = 0.9$ and $T_d = 37.2$\,K, in good agreement
with that of CYH91 who found $T_d = 37.6$\,K with $\beta$ fixed at 1.

The large extension of the AMI object makes emission from either a
protoplanetary disc or a single protostar unlikely, although the high
dust temperature suggests that this is not a prestellar core as has
been previously supposed. A spinning dust model with the dimension of
the object fixed at $\Omega_S = 30$\,arcmin$^2$, as suggested by the
AMI data would imply an $N({\rm{H}}) = (4.6\pm0.8) \times 10^{26}$\,m$^{-2}$. This spectrum is shown
in Fig.~\ref{fig:l1103spec}.

Visser et~al. (2001) comment that the distribution of material in
L1103 looks as though it may have been swept up by a wind. The
emission from objects with stellar winds and stars experiencing mass
loss has been modelled in the past (Calabretta 1991; Casassus
et~al. 2007) by treating the optical depth as a power law function of
radius. These models, parameterized correctly, can lead to a rising
flux density spectrum even into the mm-wave regime. We obtain a
spectral index of $\alpha_{16\,\rm{GHz}}^{850\,\mu\rm{m}} =
-1.14\pm0.13$, although we note that there has been no correction for
flux losses in the flux density measurements used to calculate this
index. Indeed, with no basis for a model for this source it is
not possible to adequately constrain this index in order to make any
quantitative comments. We note, however, that a radial power law model is a
vast over-simplification in this case as the emission is far from
spherically symmetric, and also that models of this type which fit
both the AMI and SCUBA data points greatly over estimate the flux
densities measured at 12 and 25\,$\mu$m. In addition, these winds are
generally associated with 
more evolved protostars. L1103 is supposedly starless, having no
observed outflows or IRAS PSC associations. However, the 12 and 25\,$\mu$m flux
densities of CYH91 would indicate a hotter dust component to this
cloud which might suggest a Class 0 object.

\begin{figure}
\centerline{\includegraphics[height=6.5cm,width=8.cm,angle=0]{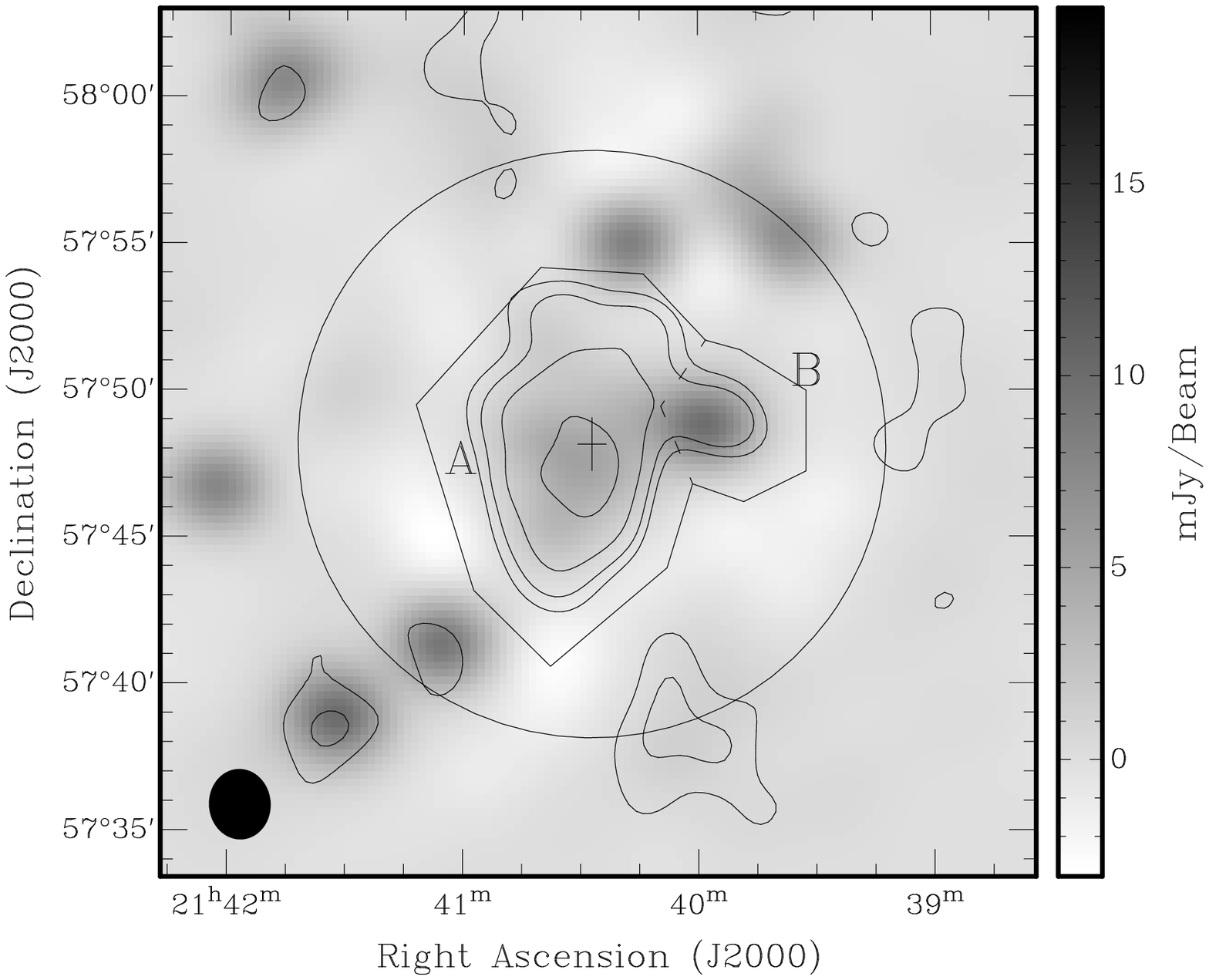}} 
\caption{L1111: CLEANed combined channel AMI data at 16\,GHz are shown as contours with the
  levels as in Fig.~\ref{fig:l675}. CLEANed {\it{uv}} matched data from the
  CGPS at 1.4\,GHz are shown
  in greyscale. The AMI pointing centre is indicated by a cross and
  the FWHM of the primary beam as a solid circle. The AMI synthesized
  beam, $2\farcm4 \times 2\farcm1$, is shown as a filled ellipse in
  the bottom left corner. Example flux
  extraction apertures are shown to illustrate the division between
  the sources labelled ``A'' and ``B''.\label{fig:l1111}
}
\end{figure}

\subsection{L1111:} The subject of a separate publication, L1111
possesses a significant excess of emission relative to that found at
1.4\,GHz in CGPS data {\it{uv}} sampled to match the AMI coverage
(AMI Consortium: Scaife et~al. 2009). Using our {\it{uv}} sampling Case (2) we also
confirm that there is an excess relative to the 2.7\,GHz flux
density. This map has significantly larger resolution
($5\farcm1\times4\farcm9$) and so the point source to the west of L1111,
denoted ``B'' in Scaife et~al., cannot be clearly resolved from the main
object. The integrated flux density of the two objects at 2.7\,GHz is
15.6$\pm1.0$\,mJy. The spectrum of source ``B'' was measured to be
$\alpha = 0.43\pm0.18$, which would give it a flux of $S_{2.7} =
7.6^{+1.5}_{-1.3}$\,mJy, leaving the object associated with L1111 a flux
density of $S_{2.7} = 8.0^{+1.3}_{-1.2}$\,mJy. However, the noise on the original Effelsberg data is very similar to
this at 
8\,mJy\,beam$^{-1}$, so 
the {\it{uv}} sampled flux density should be viewed with caution. 

Although there is no IRAS association with L1111, we fit a greybody
spectrum to the 100\,$\mu$m flux from the IRAS RSC Reject
Catalogue. Using the 100 and 850\,$\mu$m flux densities we find a best
fitting temperature of $T_d = 20$\,K using $\Omega_{\rm{S}} =
6$\,arcmin$^2$ measured from the SCUBA map. The spectral index across
the flux loss corrected AMI data is $\alpha_{14.6}^{17.2} =
-2.9\pm0.4$, see Scaife et~al. 2009 for further details.

\begin{figure}
\centerline{\includegraphics[height=8.cm,angle=-90]{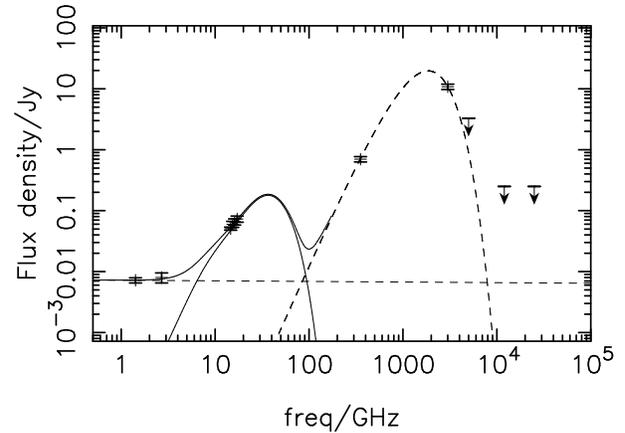}} 
\caption{SED of L1111. {\it{uv}} sampled data points are shown at 1.4
  and 2.7\,GHz from the 
  CGPS and Effelsberg 100\,m. Data at 13.8 to 17.6\,GHz
  from AMI are also shown. As are points at 850\,$\mu$m (353\,GHz) from
  Visser et~al. (2001; 2003); and at 100, 60, 25 and 12\,$\mu$m from
  the IRAS point source rejects catalogue. A power law is fitted to
  the data at 1.4 and 2.7\,GHz which is shown as a dashed line. A
  modified greybody spectrum with 
  $\beta = 1.5$ is fitted to the 850 and 100\,$\mu$m data as
  described in the text and a MC spinning dust model from DL98 is
  fitted to the AMI data points, see text for details. The combined
  SED across the AMI band is shown as a solid
  line.\label{fig:l1111spec}
}
\end{figure}

\begin{figure}
\centerline{\includegraphics[height=6.5cm,width=8.cm,angle=0]{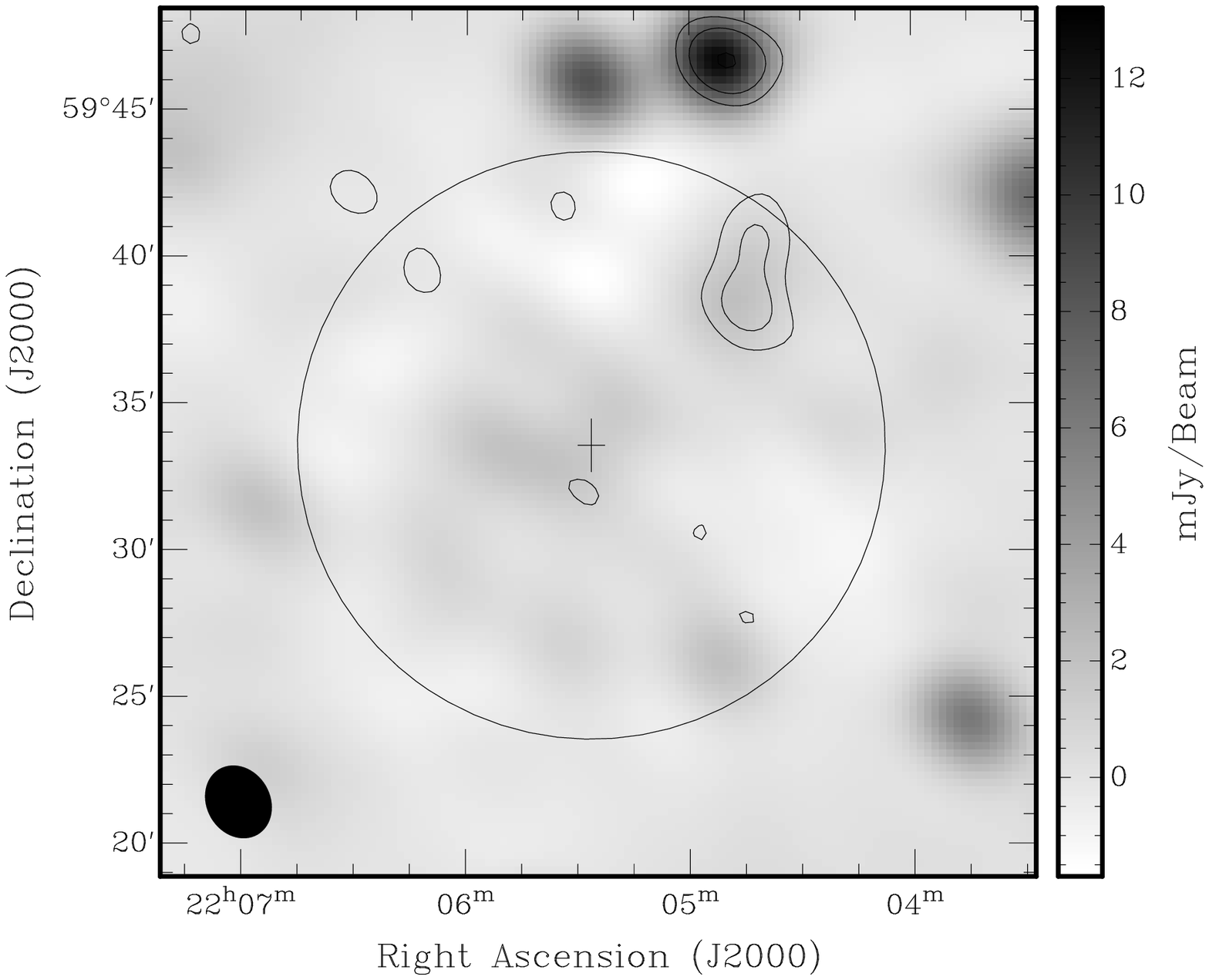}} 
\caption{L1166: CLEANed combined channel AMI data at 16\,GHz are shown as contours with the
  levels as in Fig.~\ref{fig:l675}. CLEANed {\it{uv}} matched data from the
  CGPS at 1.4\,GHz are shown
  in greyscale. The AMI pointing centre is indicated by a cross and
  the FWHM of the primary beam as a solid circle. The AMI synthesized
  beam, $2\farcm6 \times 2\farcm2$, is shown as a filled ellipse in
  the bottom left corner.  \label{fig:l1166}
}
\end{figure}

\subsection{L1166:} This object is coincident with a patch of diffuse
emission seen at 1.4\,GHz which has become significantly fainter at
16\,GHz, see Fig.~\ref{fig:l1166}.

\begin{figure}
\centerline{\includegraphics[height=6.5cm,width=8.cm,angle=0]{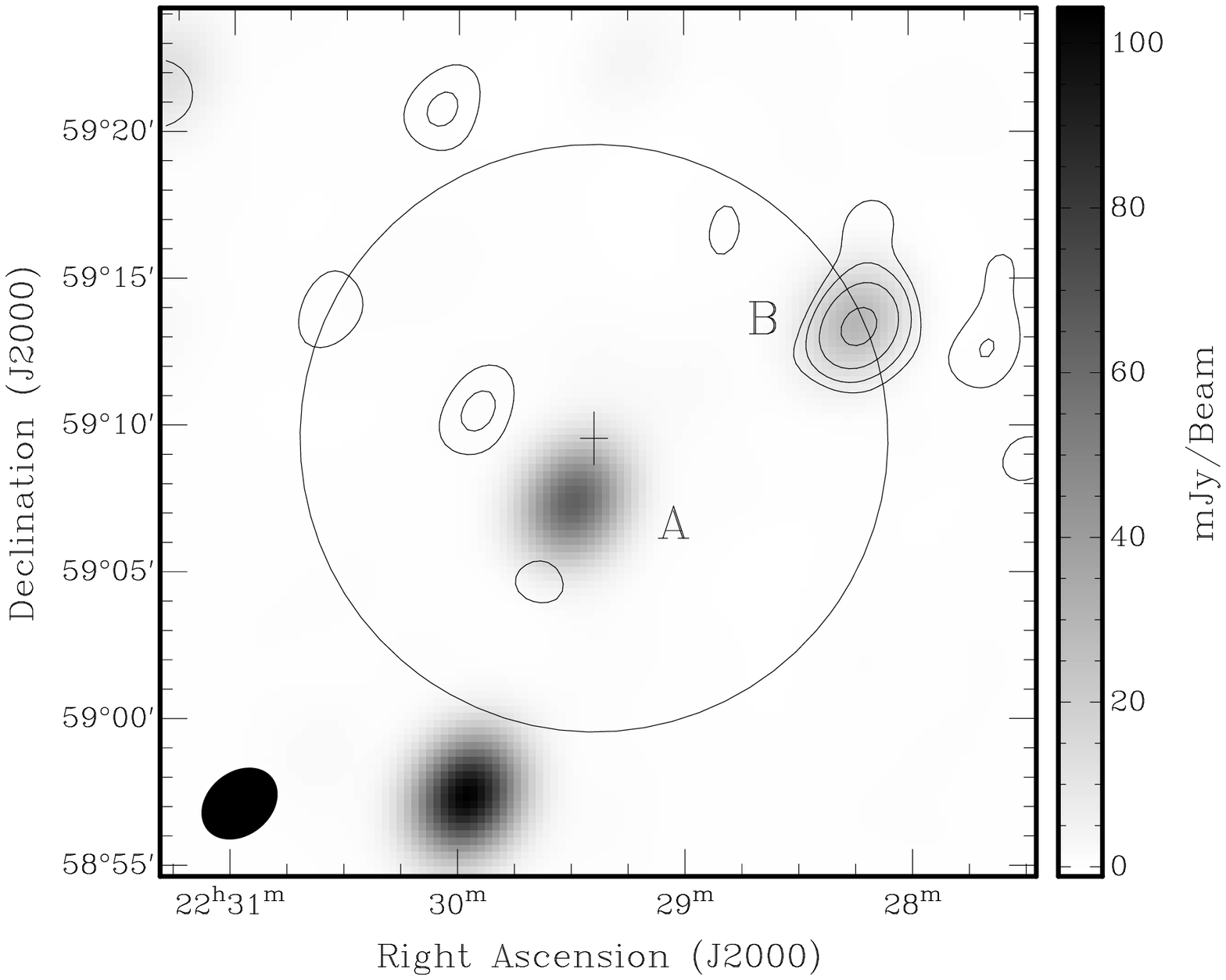}}
\caption{L1185: CLEANed combined channel AMI data at 16\,GHz are shown as contours with the
  levels as in Fig.~\ref{fig:l675}. CLEANed {\it{uv}} matched data from the
  CGPS at 1.4\,GHz are shown
  in greyscale. The AMI pointing centre is indicated by a cross and
  the FWHM of the primary beam as a solid circle. The AMI synthesized
  beam, $2\farcm8 \times 2\farcm2$, is shown as a filled ellipse in
  the bottom left corner. \label{fig:l1185}
}
\end{figure}

\begin{figure}

\centerline{\includegraphics[height=8.cm,angle=-90]{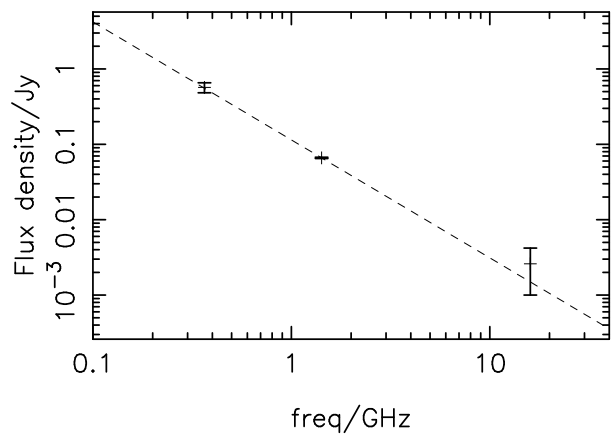}}
\caption{L1185: Radio spectrum of Source ``A''. Data points are at 365\,MHz
(Texas Survey, Douglas et~al. 1996), 1.4\,GHz ({\it{uv}} matched CGPS, this work) and 16\,GHz (AMI, this work). A
  best fitting spectral index of $\alpha = 1.3\pm1.1$ is shown as a
  dashed line.\label{fig:l1185spec}
}
\end{figure}

\subsection{L1185:} The L1185 field is relatively empty, see Fig.~\ref{fig:l1185}. The compact
source present to the south of the pointing centre in the {\it{uv}}
sampled CGPS data has a flux density of $S_{1.4} = 66.1\pm1.3$\,mJy at
1.4\,GHz and $S_{16} = 2.6\pm1.6$\,mJy ($2\sigma$) in the AMI
data. This gives it a poorly constrained spectral index of
$\alpha_{1.4}^{16}=1.3\pm1.1$, but is consistent with that
derived from data at 365\,MHz (Texas Survey) and the CGPS at 1.4\,GHz 
of $\alpha_{0.365}^{1.4} = 1.6\pm0.6$, see Fig.~\ref{fig:l1185spec}. We therefore suggest that
this is a steep spectrum extragalactic source.

\begin{figure*}
\centerline{\includegraphics[height=6.5cm,width=8.cm,angle=0]{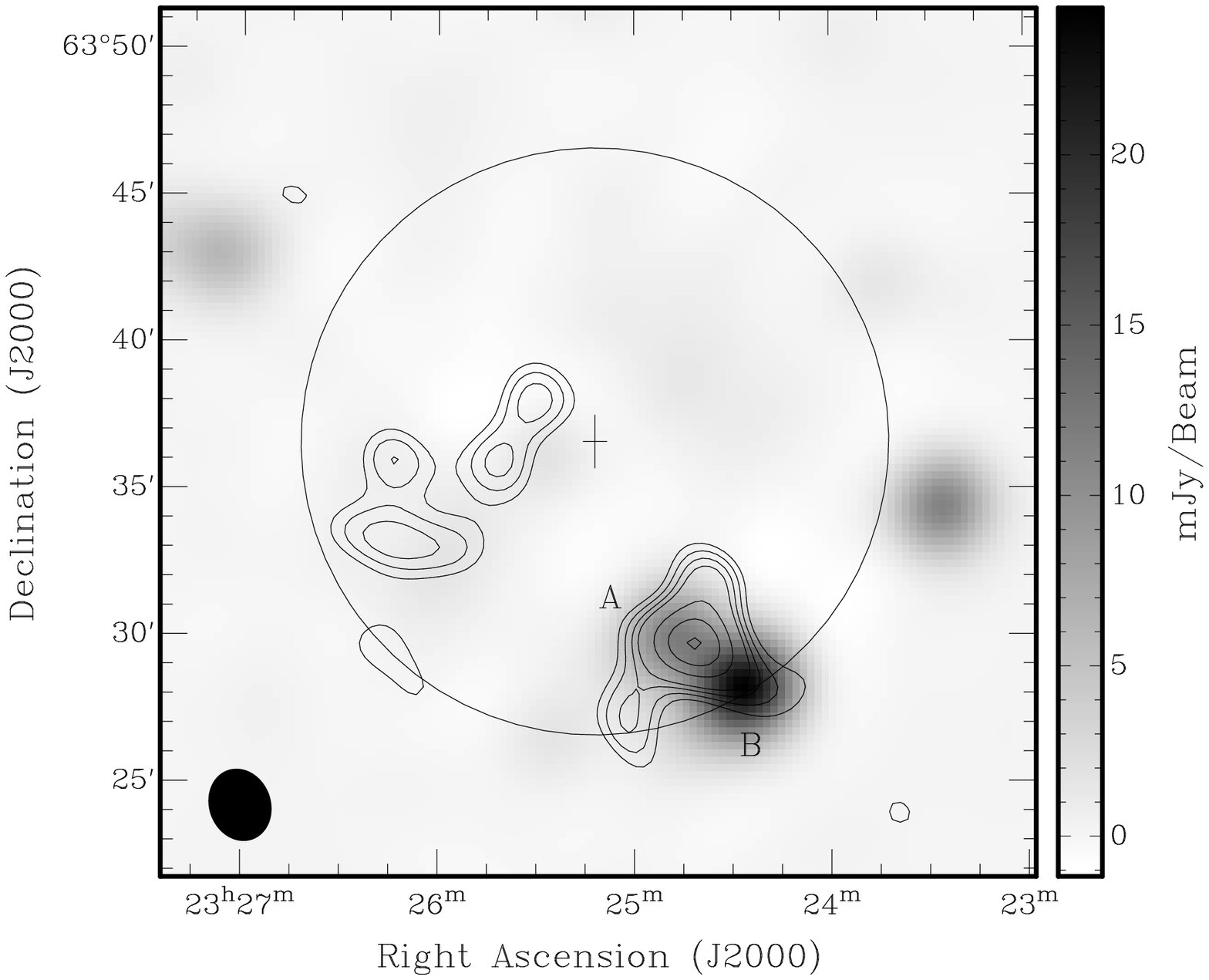}\qquad
  \includegraphics[height=6.5cm,width=8.cm,angle=0]{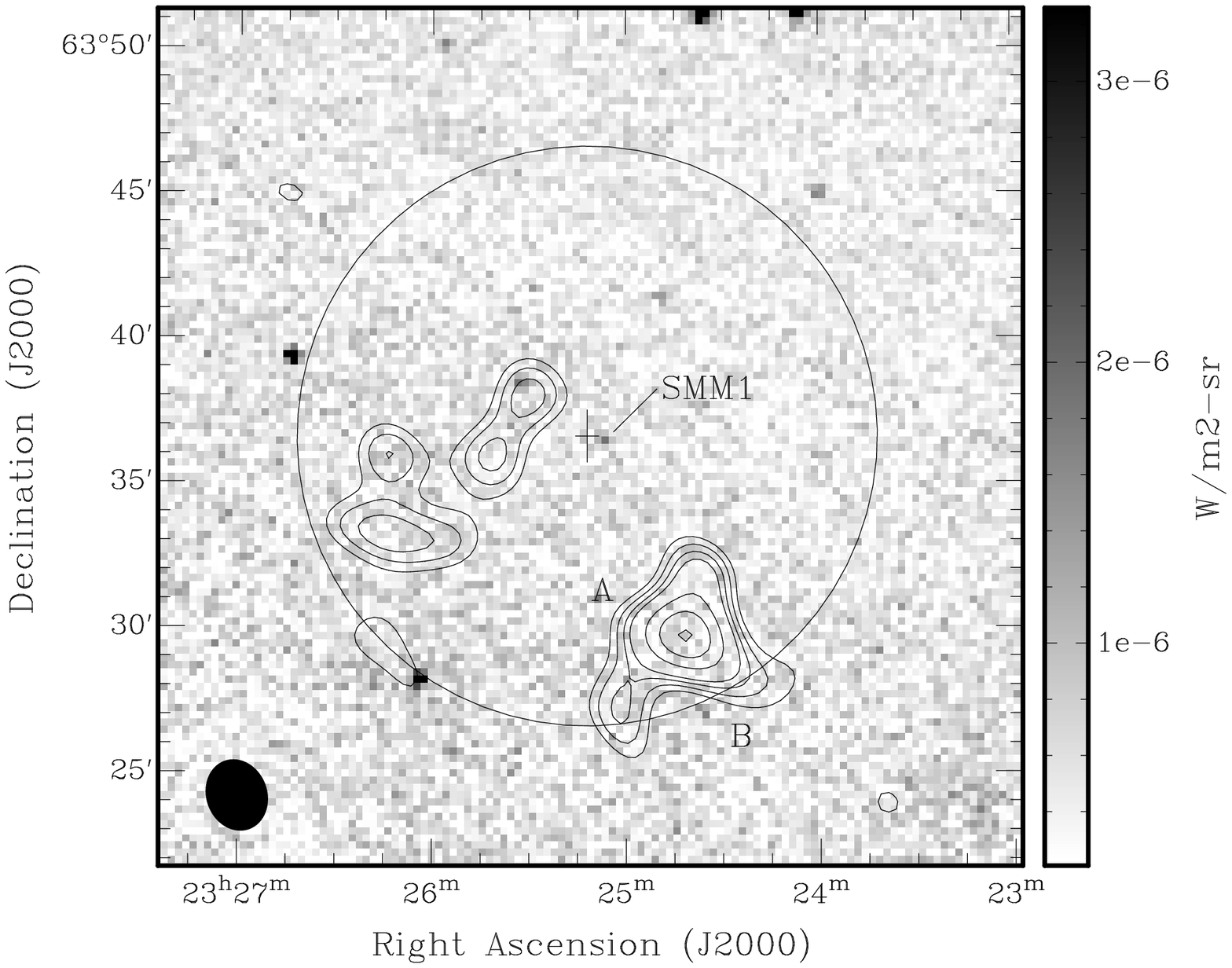}}

\centerline{(a)\hspace{8cm}(b)}

\centerline{\includegraphics[height=6.5cm,width=8.cm,angle=0]{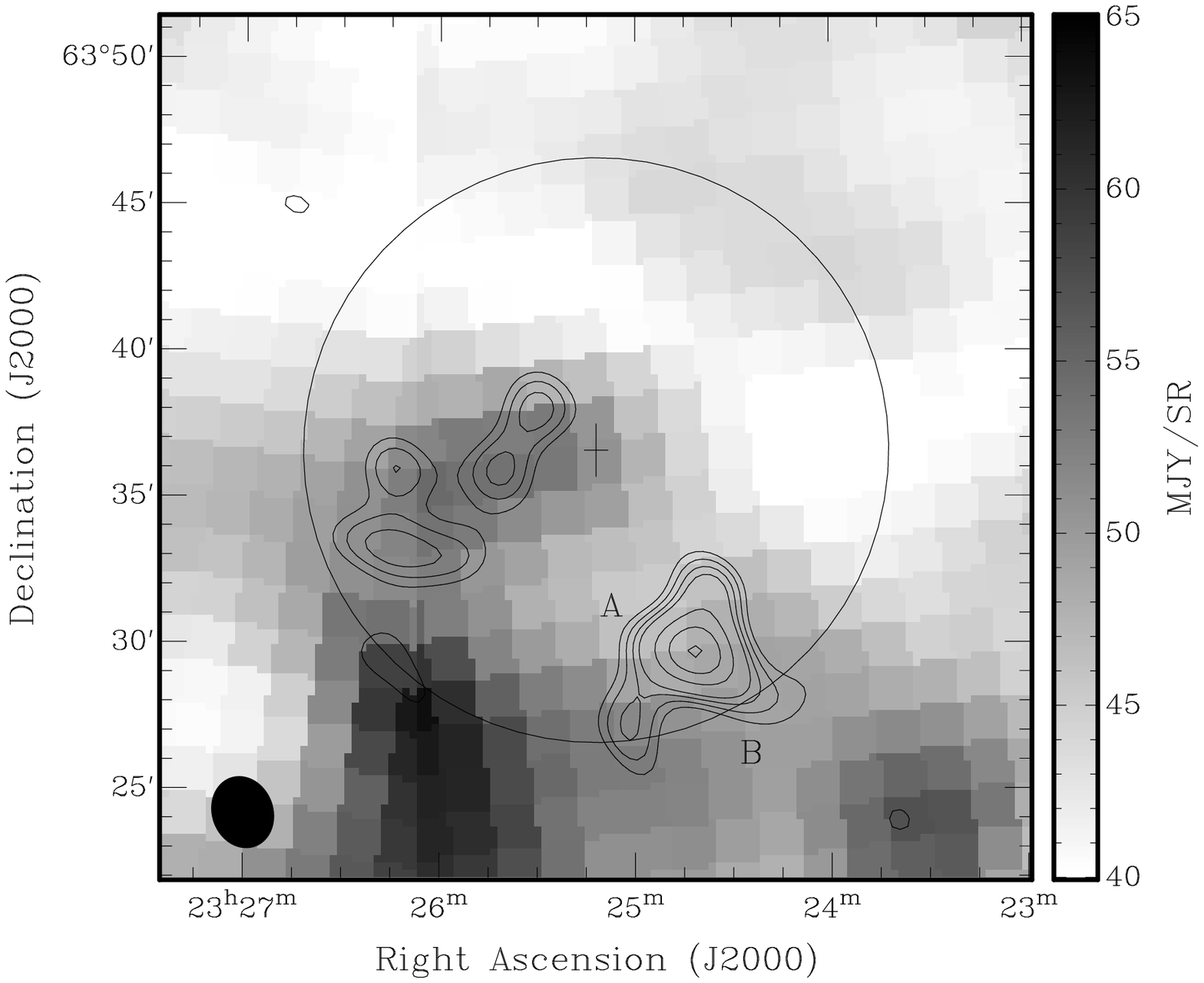}\qquad
  \includegraphics[height=6.5cm,width=8.cm,angle=0]{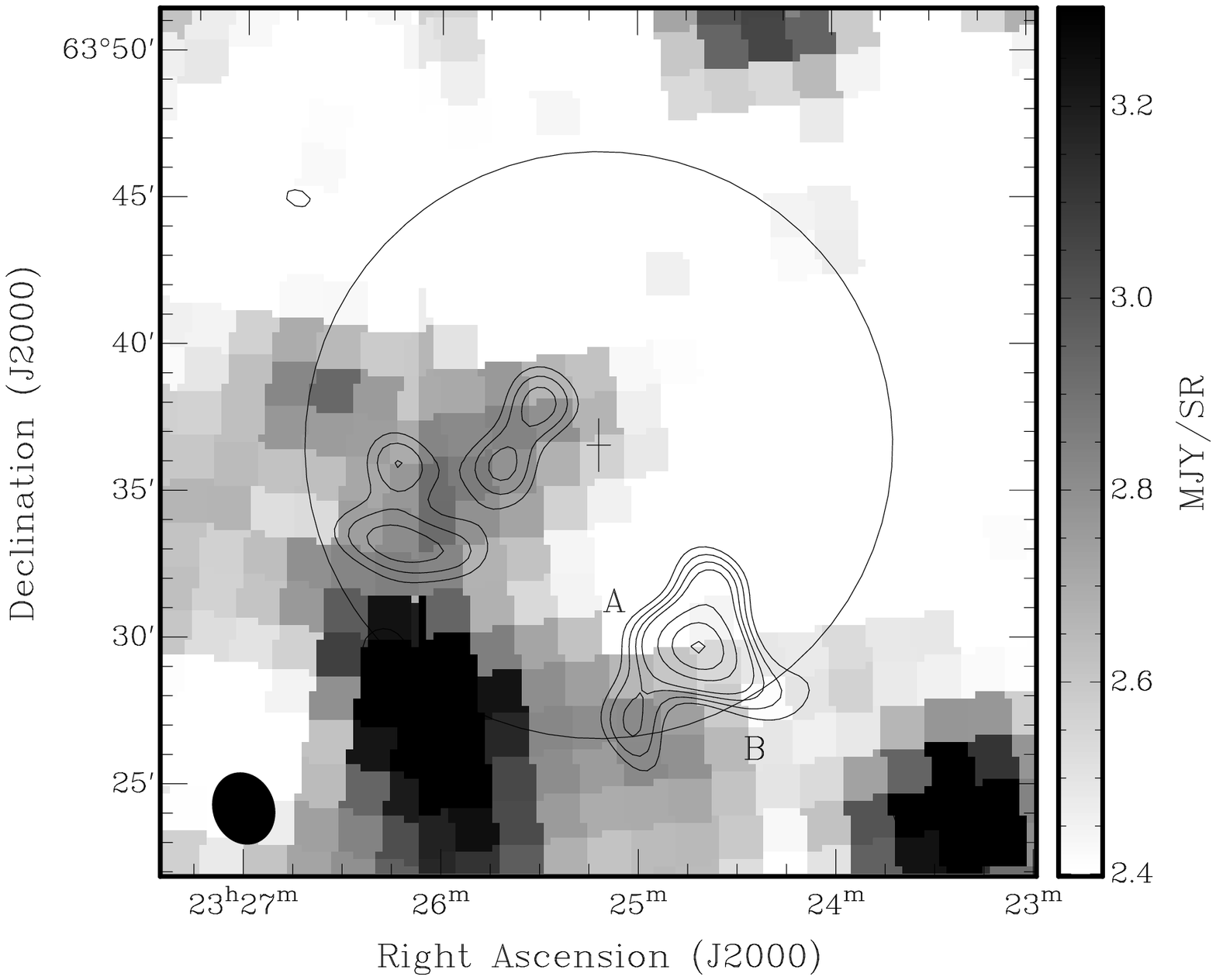}}

\centerline{(c)\hspace{8cm}(d)}

\caption{L1246: AMI 16\,GHz combined channel data are shown as contours at 3, 4, 5, 6,
  9, 12 and 15\,$\sigma$. Greyscale is (a) {\it{uv}} matched CGPS data at
  1.4\,GHz, (b) 8\,$\mu$m MSX (Band A) data, (c) 100\,$\mu$m IRAS
  data, and (d) 12\,$\mu$m IRAS data. In plot (a) the position of the
  protostar SMM1 is indicated and in plots (c) and (d) the position of
  the associated IRAS point source is shown as a square. In all
  plots the AMI pointing centre is shown as a cross and the primary
  beam as a solid circle.\label{fig:l1246}. The AMI synthesized beam, $2\farcm5 \times 2\farcm2$,
  is shown as a filled ellipse at the bottom left
  of each image. The intensity in the MSX and IRIS images is saturated
  at both ends of the scale in order to make the structure in this
  region visible.
  }
\end{figure*}

\subsection{L1246:} L1246 is a reasonably well studied dark cloud. It
is known to contain a protostar (SMM1; Launhardt et~al. 1997; Visser
et~al. 2001), IRAS 23228+6320, at 23h25m05.6s
+63$^{\circ}$36$'$34$''$.1 (J2000.0). The position of the IRAS
association seems to be dominated by the 25 and 60\,$\mu$m emission,
with the 12 and 100\,$\mu$m maps showing emission further to the
east. The AMI map at 16\,GHz shows a band of emission
which traces that found at 12\,$\mu$m, see Fig.\ref{fig:l1246}. The northern most lobe
of which is also coincident with an 8\,$\mu$m point source seen in the
MSX\footnote[1]{{\tt http://irsa.ipac.caltech.edu/applications/MSX/MSX/}} image of the same region, see Fig.~\ref{fig:l1246}. This
emission has no counterpart in the lower frequency {\it{uv}} sampled
CGPS data at 1.4\,GHz. The known protostar L1246 SMM1 can also be seen
in the 8\,$\mu$m data just to the west of the pointing centre. Two of
the four PAH mid-infrared (MIR) 
features at 7.7 and 8.6\,$\mu$m lie within MSX Band A. The unidentified MSX point source coincident with the AMI emission in
the L1246 field is only present within this band.

We note that although Visser et~al. (2001) take the size of this cloud
to be small, using the dimensions of Parker (1991) of $2\farcm8
\times 1\farcm1$, the cloud appears  larger in the DSS plates with
dimensions much closer to those measured by Clemens \& Barvainis
(1988) of $10\farcm1 \times 2\farcm2$. It is unsurprising therefore that in a
later paper Clemens et~al. (1991) measure a much larger flux density
towards this object than that listed for the IRAS association of
Parker (1988). Given the large angular extent of the region in the
co-added IRAS images of Clemens et~al. (1991) it seems likely that
these flux densities will come from an area covering the emission seen
in the AMI data. 

Previous greybody fits to L1246's SED have suggested $T_d = 19$\,K
(Launhardt et~al. 1997, Visser et~al. 2001) using data at $\lambda >
100$\,$\mu$m. The temperatures found using IRAS data alone are
higher. From the IRAS PSC the temperatures found using 100 to
60\,$\mu$m, 60 to 25\,$\mu$m and 25 to 12\,$\mu$m are 31, 66 and
101\,K respectively. From the deep IRAS photometry of CYH91 using a
much larger aperture the same temperatures are 31, 69 and
222\,K. These fits are sensitive to the angular size assumed for the
object, the discrepancy at 12\,$\mu$m suggests a significant change in
morphology at that wavelength.

The morphology of the AMI data at 16\,GHz appears to trace the
12\,$\mu$m emission and to a lesser extent the 100\,$\mu$m emission
also. Taking a 5\,arcmin square region around the emission seen in the
AMI data to the east of the pointing centre there is a slightly higher
spatial correlation between the emission seen at 12\,$\mu$m ($r=0.70$)
with that seen at 16\,GHz than there is at 100\,$\mu$m ($r=0.61$). There
is no source coincident with the AMI emission at 
either 1.4 or 2.7\,GHz. The AMI data itself towards
this source is poor due to interference and the combined channel map
is heavily weighted towards Channel 6 (16.4\,GHz) making a
determination of the spectrum using AMI data alone unreliable.

\section{Discussion and Conclusions}

Of the fourteen clouds observed here two are clear candidates for
possessing anomalous dust emission at 16\,GHz: L675 and L1111. A further three
are possible candidates: L944, L1103 and L1246. Nine of the fourteen
showed no emission inconsistent with that seen at lower radio
frequencies. 

We divide the five clouds where there is an excess of emission into
two groups on the basis of three criteria: (1) Extent of the
emission seen at 16\,GHz i.e. non-point like sources; (2) Coincident
with the sub-mm position of the cloud to within a 2$'$ radius; (3)
Likelihood of alternative explanations for the 16\,GHz emission.

We summarize our results in Table~\ref{tab:emissivity}. Column [5] of
Table~\ref{tab:emissivity} summarizes which dark clouds have a
possible spinning dust association, defined as being a microwave
counterpart within a 2\,arcmin radius which shows an excess at 16\,GHz
relative to lower frequency data. Columns [6--8] then divide the
certainty of these  
detections based on the three criteria stated in the previous
paragraph. 

On the basis of these three criteria we see that all are satisfied by
L675 and L1111 which display extended emission coincident with the
pointing centre. The extended nature of this emission makes the
possibility of the emission being cm-wave radiation from either a
protostar or proto-planetary disc unlikely. L944 shows compact
emission which is coincident with the pointing centre. The compact
nature of this emission means that cm-wave radiation from a
protostar cannot be ruled out in this case. L1246
shows emission just within a radius of 2\,arcmin from the
SCUBA position of this cloud which has no lower frequency counterpart.
However, the compact nature of this emission suggests that it may again be
cm-wave emission from a protostar.

L1103 shows extended emission which is coincident with the pointing
centre. However, although the microwave spectrum appears to be
consistent with emission from spinning dust this source has a high
flux density at 
850\,$\mu$m (Visser et~al. 2001) relative to IRAS
measurements at 12--100\,$\mu$m (CYH91). This raises questions over
the suitability of a standard greybody fit to the thermal dust
emission from this cloud and, in the absence of further observations
between 100 and 300\,GHz, it is not possible to rule out a flattened
tail to the thermal dust spectrum which may account for the excess
seen at 16\,GHz.

\begin{figure*}
\centerline{\includegraphics[height=12cm,angle=-90]{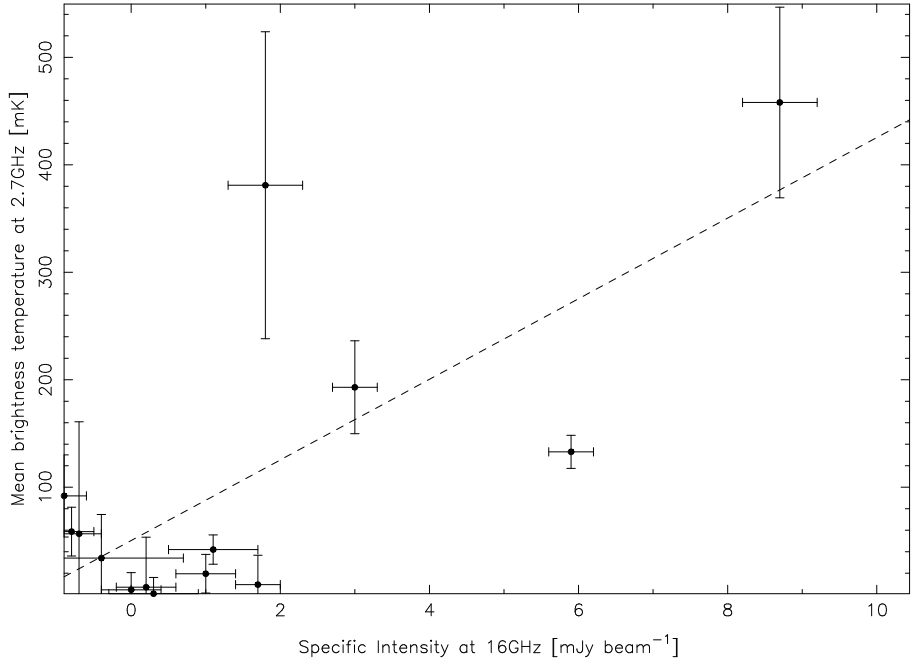}}
\caption{Correlation of mean radio brightness temperature with
  microwave specific intensity towards observation pointing
  centres. Data are specific intensity at 16\,GHz towards the SCUBA
  position of the dark cloud on the x-axis, and mean brightness temperature
  within a 1$^{\circ}$ diameter circle centred on the position of the
  dark cloud on the y-axis. The error bars on this mean show the
  standard deviation of the brightness temperature distribution within
  this region.\label{fig:11cmcor}}
\end{figure*}

In their theoretical models for the emission from rotating grains DL98
considered a number of gas-grain, plasma, and radiative contributions to the
excitation and damping of spinning dust grains. For their several
models of grain environment it was found that collisions with ions
were the dominant excitation mechanism in all but two scenarios:
reflection nebulae and photodissociation regions, where photoelectric
effects dominated due to the intense radiation field. For the
environments relevant to dark nebulae (molecular cloud and cold
neutral medium environments) this suggests
that objects which are embedded in or exposed to strongly 
ionized environments are the most
likely candidates for observing spinning dust emission. The presence
of an anomalous component in the spectra of the objects presented here 
appears to be correlated with the average free-free emission within a
1 degree area centred on each pointing centre, see Fig~\ref{fig:11cmcor}. 
We assess this correlation following the method of Franzen
et~al. (2009), where we express the covariance matrix of our data
points, $C$, as

\begin{eqnarray}
C&=&S+N_i\\
 &=& \left( \begin{array}{cc}
S_{11} & S_{12} \\
S_{21} & S_{22} \end{array} \right) + \left( \begin{array}{cc}
\sigma_{x,i}^2 & 0 \\
0 & \sigma_{y,i}^2 \end{array} \right).
\end{eqnarray}

$S_{11}$ and $S_{22}$ are the variances of the 2.7 and 16\,GHz data
points respectively and $S_{12} (=S_{21})$ is the covariance of
the 2.7 and 16\,GHz data points. It follows that the correlation
co-efficient, denoted $R$ to distinguish it from the standard Pearson co-efficient, is given by

\begin{equation}
R = \frac{S_{12}}{\sqrt{S_{11}S_{22}}}.\label{equ:rcoeff}
\end{equation}

Following Franzen et~al. (2009) we used a Gaussian likelihood and the nested
sampling algorithm implemented in the MultiNest code (Feroz \& Hobson
2008; Feroz, Hobson \& Bridges 2008) to obtain $R$. For the averaged
2.7\,GHz flux density we find a correlation coefficient of
$R=0.81\pm0.15$. For the 2.7\,GHz flux density at the position
of the cloud with the 16\,GHz flux density at the same position we
find a weaker correlation, $R=0.66\pm0.20$. We repeated these
evaluations setting the noise covariance matrix to zero and found
values of $R=0.70\pm0.18$ and $R=0.65\pm0.20$ for the averaged
flux density at 2.7\,GHz and the 2.7\,GHz flux density at the position
of the cloud, respectively. We can see that the noise in the first
instance leads to an decrease in correlation unless it is accounted
for. 

Of the fourteen clouds observed, those with the highest background
levels at 2.7\,GHz are L675, L1111, L860 and L1103; all of which have
a background level above 100\,mK. Indeed the correlation is dominated
by the high surface brightness regions, suggesting a radio surface
brightness threshold between those regions which exhibit spinning
dust emission and those which do not.

There appears to be no correlation between 16\,GHz
specific intensity and 100\,$\mu$m emission. Following the same
correlation analysis a described in Equ.~\ref{equ:rcoeff} we obtain a
value of $R=0.02\pm0.32$ ($R=-0.01\pm0.32$ without noise). The
dust emissivity at 16\,GHz 
relative to the IRIS 100\,$\mu$m data is tabulated in
Table~\ref{tab:emissivity}. We also list the predicted specific
intensities for these objects for a relative emissivity of 10
$\mu$K\,(MJy\,/sr)$^{-1}$ for reference. The relative emissivity of
our two strongest 
detections is similar to that found for LDN\,1622 (Casassus
et~al. 2006) and G159.618.5 (Watson et~al. 2005). This suggests that
the non-detections in our sample are not an effect of our sensitivity
limits. If the emissivity of anomalous microwave emission relative to
that seen at 100\,$\mu$m was at a
similar level in all dark clouds, for example $>10\,\mu$K\,(MJy/sr)$^{-1}$,
then we would have detected emission from every cloud in our sample at
$>3\,\sigma_{\rm{rms}}$. The weighted average dust 
emissivity relative to the 100\,$\mu$m maps for our sample is
6.6$\pm$5.0\,$\mu$K\,(MJy\,/sr)$^{-1}$. This value agrees with the
all-sky WMAP value for cool dust clouds at high galactic latitudes
(Davies et~al. 2006), given as 1\,Jy at 30\,GHz per 6000\,Jy at 100\,$\mu$m.

\begin{table*}
\begin{center}
\caption{16\,GHz dust emissivities relative to 100\,$\mu$m
  emission. Column [1] Source name, Column [2] Predicted intensity
  based on an emissivity of 10\,$\mu$K\,(MJy/sr)$^{-1}$, Column [3]
  measured intensity towards phase centre, Column [4] Measured 16\,GHz
  dust emissivity relative to IRIS 100\,$\mu$m data, Column [5]
  Possible identification of a spinning dust counterpart as described
  in Section 5, Column [6--8] Certainty criteria as described in
  Section 5. \label{tab:emissivity}}
\begin{tabular}{lccccccc}
\hline
\hline
Name&$S_{\rm{pred}}$&$S_{\rm{act}}$&$\epsilon$&Possible&\multicolumn{3}{c}{Criterion:}\\
&(mJy beam$^{-1}$)&(mJy beam$^{-1}$)&$\mu$K\,(MJy/sr)$^{-1}$&Identification&[1]&[2]&[3]\\
\hline
{\textbf{L675}} & 3.6&5.9$\pm$0.3&16.4$\pm$1.8&y&y&y&n\\
L709 &2.8&$-0.8$$\pm$0.3&$<0$&n&-&-&-\\
L860 &4.2&3.0$\pm$0.3&7.2$\pm$1.0&n&-&-&-\\
L917 & 7.9&$-0.7$$\pm$0.3&$<0$&n&-&-&-\\
{\textbf{L944}} & 2.7&1.7$\pm$0.3&6.2$\pm$1.3&y&n&y&n\\
L951 & 1.9&$-0.9$$\pm$0.3&$<0$&n&-&-&-\\
L953 & 2.4&1.1$\pm$0.6&4.5$\pm$2.5&n&-&-&-\\
L1014 & 2.5&0.0$\pm$0.4&0.0&n&-&-&-\\
L1021 & 2.8&1.0$\pm$0.4&3.5$\pm$1.4&n&-&-&-\\
{\textbf{L1103}} & 4.7&1.8$\pm$0.5&3.9$\pm$1.2&y&y&y&y\\
{\textbf{L1111}} & 3.1&8.7$\pm$0.5&28.5$\pm$3.3&y&y&y&n\\
L1166 & 1.5&0.3$\pm$0.6&2.0$\pm$4.0&n&-&-&-\\
L1185 & 4.2&$-0.4$$\pm$1.1&$<0$&n&-&-&-\\
{\textbf{L1246}} & 2.1&0.2$\pm$0.4&1.0$\pm$2.0&y&n&y&n\\
\hline
\end{tabular}
\end{center}
\end{table*}

Of our five possibly anomalous objects two are known to be Class 0
objects: L944 and L1246. High sensitivity maps in CO towards our two
most anomalous 
objects do not yet exist and they are assumed starless. However the
starless/protostellar divide has 
come under scrutiny recently following the discovery of
Very Low Luminosity Objects (VeLLOs) with \emph{Spitzer}. VeLLOs are faint
infrared point sources with protostellar colours towards objects
previously classified as starless. Young et~al.\ (2004) discovered a
VeLLO in L1014 (L1014-IRS), part of the Visser et al.\
sample, coincident with the peak of dust continuum emission that had
been previously classified as starless by Parker (1988). Although here
no anomalous microwave emission is seen towards L1014 it is suggestive
that the peak of the anomalous emission towards 
L944 is more coincident with the centre of the red-shifted CO $2\to 1$
outflow lobe, 
mapped by Visser et al., and not the compact SCUBA emission
from the Class 0 protostar. This spatial
correspondence immediately raises the question of whether the outflow
itself or possibly the cloud turbulence could be energising the
spinning dust emission. Regardless of this possible connection we also
point out that the 
misidentification of radio emission from protostars is a potential source
of confusion when searching for emission from spinning dust. 

There seems to be no obvious correlation between the AMI flux
densities and the IRAS or SCUBA flux densities, however the number of
data points is low. It is also true that there is no correlation
between the SCUBA flux densities at 850\,$\mu$m and those of IRAS at
100\,$\mu$m. This is perhaps not particularly surprising in light of
the mismatch in measured angular scales, and would suggest that the
morphology of these objects is larger at 850\,$\mu$m than the SCUBA
chop allows for. Obviously this is not a problem when investigating
compact cores as in Visser et~al. (2001; 2002), however it provides
little assistance when looking at the correlation between cm-wave amd
mm-wave emission.

In conclusion, we have observed a sample of fourteen compact 
Lynds dark nebulae. We have found a significant excess towards two of
the fourteen and an indication of anomalous behaviour in three further
clouds. We suggest that the excess we see is due to rotational
emission from very small grains and that this emission may be
correlated with a high level of background emission at lower radio frequencies.

\section{ACKNOWLEDGEMENTS} 

We thank the staff of the Lord's Bridge observatory for their
invaluable assistance in the commissioning and operation of the
Arcminute Microkelvin Imager. We thank  John Richer
for useful 
discussions. We thank the anonymous referee, whose comments have
significantly improved this paper. The AMI is supported by Cambridge
University and the  
STFC. NHW, MLD, TF, CRG  and TS  
acknowledge the support of PPARC/STFC studentships.

\bsp
\label{lastpage}

\end{document}